\documentclass[journal=jpcafh,manuscript=article]{achemso}
\pdfoutput=1
\usepackage[T1]{fontenc} 
\usepackage{thinsp}
\usepackage{booktabs}
\usepackage{array}
\usepackage{listings}
\usepackage{lmodern}
\usepackage{microtype}
\usepackage{caption}
\usepackage{float}
\usepackage{setspace}
\usepackage{xkeyval}

\usepackage{amsmath}
\usepackage{amssymb}
\usepackage{graphicx}
\usepackage{bm}
\usepackage{color}
\usepackage{verbatim}
\usepackage{amssymb}
\usepackage{fixltx2e}

\newcommand{\bra}[1]{\langle #1 |}
\newcommand{\ket}[1]{| #1 \rangle}

\newcommand{\ep}{\epsilon}
\newcommand{\no}{\nonumber}
\newcommand{\eqr}[1]{Eq.~\eqref{eq:#1}}

\newcommand{\eql}[1]{\label{eq:#1}}
\newcommand{\figr}[1]{Fig.~\ref{fig:#1}}
\newcommand{\figl}[1]{\label{fig:#1}}
\newcommand{\bse}{\begin{subequations}}
\newcommand{\ese}{\end{subequations}}
\newcommand{\mM}{\mathcal{M}}
\newcommand{\mcN}{\mathcal{N}}

\newcommand{\rCT}{\mathrm{CT}}
\newcommand{\rLE}{\mathrm{LE}}

\SectionNumbersOn
\author{Timothy J.\ H.\ Hele}
\email{tjhh2@cam.ac.uk}
\affiliation{Cavendish Laboratory, JJ Thomson Avenue, Cambridge University, CB3 0HE, UK}
\author{Eric G. Fuemmeler}
\affiliation{Baker Laboratory, 259 East Avenue, Cornell University, Ithaca, NY 14850, USA}
\author{Samuel N. Sanders}
\author{Elango Kumarasamy}
\affiliation{Department of Chemistry, Columbia University, New York, NY 10027, USA}
\author{Matthew Y. Sfeir}
\affiliation{Center for Functional Nanomaterials, Brookhaven National Laboratory, Upton, NY 11973, USA}
\author{Luis M. Campos}
\email{lcampos@columbia.edu}
\affiliation{Department of Chemistry, Columbia University, New York, NY 10027, USA}
\author{Nandini Ananth}
\email{ananth@cornell.edu}
\affiliation{Baker Laboratory, 259 East Avenue, Cornell University, Ithaca, NY 14850, USA}

\title{Anticipating acene-based chromophore spectra with molecular orbital arguments}

\abbreviations{}

\begin{document}

\begin{abstract}
%
Recent synthetic studies on the organic molecules tetracene and pentacene have
found certain dimers and oligomers to exhibit an intense absorption in the 
visible region of the spectrum which is not present in the monomer or 
many previously-studied dimers. In this article we combine experimental 
synthesis with electronic structure theory and spectral computation to 
show that this absorption arises from an otherwise dark charge-transfer excitation `borrowing intensity' from an intense UV excitation. 
Further, by characterizing the role of relevant monomer molecular orbitals, 
we arrive at a design principle that allows us to predict the presence or 
absence of an additional absorption based on the bonding geometry of the dimer. 
We find this rule correctly explains the spectra of a wide range
of acene derivatives and solves an unexplained structure-spectrum phenomenon first
observed seventy years ago. 
These results pave the way for the design of highly absorbent
chromophores with applications ranging from photovoltaics to liquid crystals.
\end{abstract}


\section{Introduction}
Investigation of the electronic structure and spectra of organic molecules has a long history, 
dating back to the molecular orbital theories of H\"uckel\cite{huc30a} and the oscillator strength 
sum rules of Thomas, Reiche and Kuhn (TRK)\cite{tho25a,rei25a,kuh25a}. 
These have been followed by development and application of more sophisticated electronic 
structure methods\cite{pop53a,pop55a,par53a,par56a,zen14a,haj09a,cas09a,son07a,has96a,kaw99a} 
and the formulation of structure-spectrum design principles such as Kasha's 
point-dipole model\cite{kas65a}, crystallochromy\cite{kaz94a}, and the tuning of 
absorption frequencies (color) by alteration of orbital energy gaps\cite{han10a}. 
Despite this considerable progress, there remains a need for more systematic design rules 
for the creation of absorbent and tunable chromophores\cite{mis09a,zol03a,chr15a}.

More recently, there has been a surge of interest in organic chromophores for the 
development of efficient  photovoltaics\cite{kip09a,for05a}. 
In particular, acenes such as tetracene and pentacene possess the unusual 
and useful ability to undergo singlet fission\cite{cha11a,wal13a,san15a,zir15a} 
which, by splitting one high-energy exciton into two low-energy triplets, 
has the potential to substantially increase the efficiency of organic 
solar cells.\cite{smi10a,smi13a} 
This has led to the synthesis and characterization of a large range of 
dimers and oligomers of tetracene and pentacene, some of which exhibit 
an intense visible absorption in addition to the monomer $S_0 \to S_1$ 
transition\cite{pun17a,san15a,san16b}, whereas others do not\cite{luk15a}. 
This additional absorption occurs without any significant change in the 
intensity or frequency of the lowest-energy monomer ($S_0\to S_1$) excitation, 
which is preserved in all these dimers\cite{pun17a,san15a,san16b,luk15a}. 
Examining the literature, we find that an interesting dependence of UV-vis
spectra on the bonding geometry of acene dimers was first observed in 1948 for 
dimers of naphthalene (where all the transitions are in the UV),\cite{fri48a} 
and has since been observed in many other acene derivatives \cite{yam16a,rud14a,gey15a,bal11a}. 
Previous computational investigations have assigned this to 
a $\pi \to \pi^*$ transition \cite{yam16a,kum16a} but there has been no clear 
explanation of its presence in some acene derivatives \cite{fri48a,yam16a,rud14a,gey15a,bal11a} 
and absence in others \cite{leh10a,bar11a,aki60a,lon01a,bho13a,coo16a,bul13a,gu15a}.

While initially a curiosity, the ability to synthesize acene derivatives that exhibit 
enhanced visible absorptions may lead to organic photovoltaic materials with increased efficiency.
In theory, using a more absorbent chromophore would allow for a decrease in the 
thickness of a cell while still absorbing the same proportion of solar radiation 
thereby reducing cost and increasing flexibility. 
The thinner cell would also allow more diffusing excitons to be extracted as current, 
increasing efficiency\cite{for05a,zhe16a}. There is, consequently, a theoretical 
and practical need to uncover the origin of this new absorption for the design 
of novel acene-derivative based materials.

In this article we present a joint experimental-theoretical investigation
that finds the origin of this new visible absorption is `intensity borrowing'\cite{rob67a} 
from an intense UV absorption. 
The presence or absence of intensity borrowing in dimers can be deduced by 
examining whether the relevant monomer molecular orbitals have amplitude 
on the carbon atoms through which the monomers are bonded. 
We find that the resulting design rule can be used to predict the absorption 
spectra of a very wide range of acene derivatives, including oligomers
with unusual bonding geometries and heteroatom-substituted derivatives. 

We begin by presenting experimental and computed acene spectra 
in section~\ref{sec:exp}, finding that Pople-Parr-Pariser (PPP)
theory\cite{pop53a,pop55a,par53a,par56a} 
correctly predicts the presence or absence of the extra absorption 
and its approximate intensity. 
To explain why this additional absorption is observed in some dimers
and not others, in section~\ref{sec:intbor}, we use intensity borrowing 
perturbation theory and PPP theory to analyze 
the relevant monomer molecular orbitals.
In section~\ref{sec:molorb} we apply this analysis to bipentacenes, explaining 
the origin of the new absorption in some dimers and its absence in others,
and we formulate a general design rule for acene derivatives. 
We show that this design rule correctly predicts the presence 
or absence of an additional absorption in a large and diverse 
range of dimers, trimers, and oligomers in section~\ref{sec:app}.
Conclusions are presented in section~\ref{sec:conc}.

\section{Acene dimer spectra}
\label{sec:exp}
The spectra of acenes is a much-studied area 
(see, for example,\ Refs.~\citenum{cou48a,par56a,dew54a,haj09a,aki60a,aki60b,tan65a,cla79a,has96a,kaw99a,hal00a,pay05a,leh08a,haj09a,yam11a,son07a,bel13a,kur14a,zen14a}) 
and here we focus on the particular case of covalently-linked dimers and 
oligomers of acenes. 
Although we find our results to be widely applicable, we present them in 
the context of Bis(triisopropylsilylethynyl)pentacene 
(TIPS-pentacene, Fig.~\figr{struc}A), a molecule of interest for its 
extensive applications in photovoltaics~\cite{wal13a,kum16a,san15a,fue16a,leh10a}. 

\subsection{Experimental spectra}
We start by comparing the experimental spectra of TIPS-pentacene 
and a recently-reported dimer, 2,2-bipentacene (22BP)\cite{san15a,fue16a}, 
with molecules shown in \figr{struc} and spectra in \figr{spec}. 
The spectra of TIPS-pentacene and 22BP have the familiar features of 
an acene spectrum\cite{par56a,son07a,hal00a}: a weak $y$-polarized 
transition at low frequencies (coordinates defined in \figr{struc}A), 
accompanied by a vibrational stretching progression and a very intense 
absorption in the near UV (310-350~nm). We note two distinct differences
between the spectrum of 22BP and the monomer. First, 22BP exhibits 
a visible absorption around 500nm with its own vibrational stretching 
progression. Second, although the lowest energy transition 
($S_0 \rightarrow S_1$) is unaffected by dimerization, the intensity
of the UV absorption decreases and red-shifts.
The extinction coefficients in \figr{spec} are plotted 
per pentacene to show that, in the visible, the 2,$2'$ dimer 
is more absorbent than the sum of its parts and not simply 
an artifact of the chromophore being twice as large. 

\begin{figure}[tb]
 \includegraphics[width=.9\textwidth]{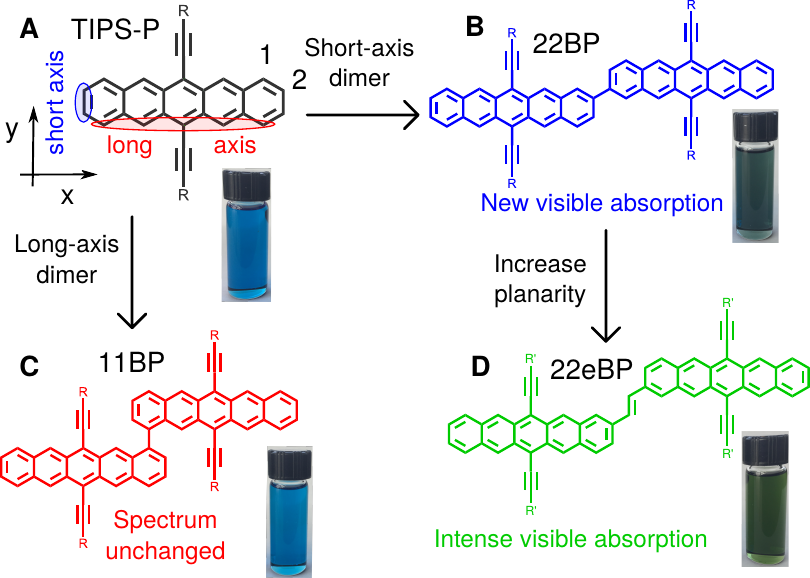}
 \caption{Structures of TIPS-pentacene (\textbf{A}), 2,2$'$-bipentacene (22BP, \textbf{B}), 1,1$'$-bipentacene (11BP, \textbf{C}) and 2,2$'$-alkene-linked bipentacene (22eBP, \textbf{D}). On TIPS-pentacene the coordinate axes are given along with definitions of short-axis (cata) and long-axis (peri) positions. By each molecule is a picture of an experimental solution in CHCl$_3$, showing how both TIPS-P and 11BP are light blue, 22BP is blue-green, and 22eBP is dark green. The corresponding linear absorption spectra are in \figr{spec}.}
 \figl{struc}
\end{figure}

\begin{figure}[tb!]
 \includegraphics[width=.5\textwidth]{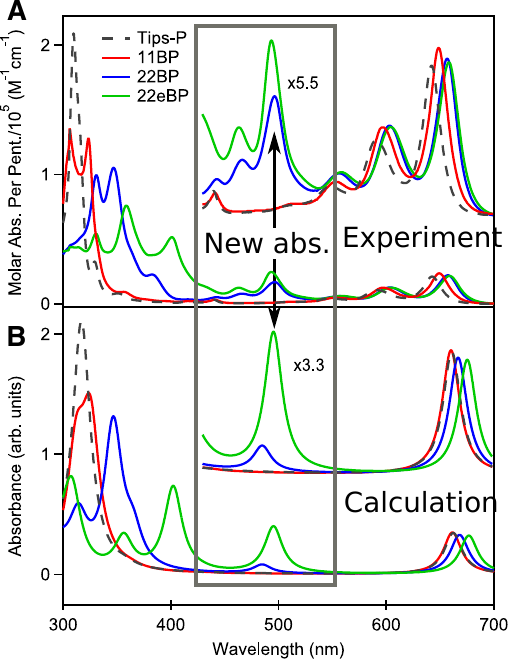}
 \caption{Experimental spectra of TIPS-P (black dashes), 11BP (red), 22BP (blue), and 22eBP (green) taken in CHCl$_3$ (\textbf{A}) and corresponding computed spectra (using PPP theory) (\textbf{B}), with insets in the visible region. Although the low-energy transition around 650nm is virtually unchanged upon dimerization, the yellow box highlights the emergence of a new absorption around 500nm, seen in 22BP and 22eBP (where it is more intense) but not 11BP. The experimental extinction coefficients and computed spectra are plotted per pentacene to show that 22BP and 22eBP are more absorbent in the visible than the sum of their parts. Computed spectra have linebroadening applied but have not been shifted to agree with experiment and do not incude vibrational progressions.}
 \figl{spec}
\end{figure}

A previous study focusing on the singlet fission properties of these 
molecules\cite{fue16a} considered bonding bulking phenyl groups around 
the inter-pentacene bond, finding that the size of the new absorption 
decreased as the twist angle between the two chromophores increased. 
Similarly, bonding a succession of benzene groups as linkers between 
the two pentacenes causes the new absorption to decrease in intensity.
To investigate the origin of this visible absorption further, we synthesized 
two more pentacene derivatives (details in the supporting information). 
We make 1,1'-bipentacene (11BP, \figr{struc}C) finding its spectrum in 
\figr{spec} to be qualitatively identical to the TIPS-P monomer spectrum 
in the visible, and with a splitting of the high-energy absorption 
in the UV. We also synthesize a planar analogue of 22BP, 2,2'-ethene-linked 
bipentacene (22eBP, \figr{struc}D) by linking the monomers with an 
alkene bond, where the alkene bond can be loosely interpreted as a 
conjugated `wire' between two chromophores. We find that 22eBP exhibits
the most intense absorption in the visible region of the spectrum 
among the bipentacenes considered here.

\subsection{Computed spectra}
\label{sec:ppp}
Since these are very large chromophores, we employ Pariser-Parr-Pople (PPP) 
theory~\cite{par56a,pop53a,suz73a,kou67a,pac06a} to calculate dimer spectra. 
PPP theory was developed for the calculation of low-lying excited electronic 
states of conjugated systems\cite{par56a,pop53a}, and its similarity to 
simple molecular orbital (MO) theories such as H\"uckel theory allows for a 
chemically intuitive interpretation of results \cite{suz73a,kou67a}. 
Unlike H\"uckel theory, however, PPP theory includes two-electron interactions 
(within the neglect of differential overlap (NDO) approximation) such that it 
correctly accounts for Coulomb and exchange interactions which are required to 
give accurate excited-state energies \cite{par53a}. 
Computational details are given in the SI where we also benchmark our monomer 
calculations against high-level multiconfiguational calculations, confirming 
the suitability of PPP theory.

Examining \figr{spec}, we see that the PPP spectra are in good agreement with experimental spectra (the computed energies are not shifted to agree with experiment) and correctly predicts a visible absorption in 22BP and 22eBP that is not present in either 11BP or the monomer. While the visible absorption in 22BP is somewhat weaker and slightly blue-shifted compared to experiment, in 22eBP it is accurately captured, both in terms of absorption frequency and relative intensity. We also see that PPP correctly predicts a Davydov splitting of the 11BP absorption (seen as a peak with a shoulder around 320nm), and a diminution in the height of the UV peaks of 22BP and 22eBP, although their relative intensities are inaccurate.

From our PPP  calculations (verified by high-level calculations reported in the SI), 
we attribute the dimer excitation around 650nm to a predominantly HOMO$\to$LUMO transition. 
The intense UV peak arises from an in-phase combination of HOMO--3 to LUMO and HOMO to LUMO+3, whose corresponding out-of-phase combination is dark in PPP theory and is seen as a weak absorption at 440~nm in experiment. Excitations from the HOMO--1 to LUMO and HOMO to LUMO+1 are dipole forbidden, and the excitations from the HOMO--2 to the LUMO and the HOMO to the LUMO+2 are predicted by PPP theory to be $y$-polarized, giving a dark out-of-phase state and a bright in-phase state at 358nm and 366nm respectively. The bright state probably corresponds to a shoulder in the intense UV absorption seen experimentally around 350nm.

\section{Elucidating dimer spectra}
\label{sec:intbor}
We investigate the origin of the visible absorption in 22BP and 22eBP using 
intensity borrowing perturbation theory\cite{rob67a}, where we contruct a 
zeroth-order set of states (corresponding to non-interacting monomers), 
and examine the perturbation (if any) introduced by dimerization by bonding
at different positions.

We begin by noting that textbook arguments\cite{atk11a} fail to explain 
the variation in our dimer spectra. Conventional symmetry-breaking arguments 
fail to explain the extra absorption since both 11BP and 22BP are $C_2$ 
and the more symmetric $C_{2h}$ 22eBP has the largest absorption (see \figr{spec}). 
Similarly, a particle-in-a-box model\cite{atk11a} predicts an intense 
HOMO to LUMO $S_1$ transition, which is inaccurate even for the monomer 
(\figr{spec}), and it also predicts that the lowest-energy transition 
redshifts and grows in intensity upon enlarging the molecule, 
which is not observed. 
Furthermore, solvatochromism cannot explain the extra absorption, 
since alternant hydrocarbons\cite{cou40a} such as acenes have no 
ground-state dipole or multipole. 
Planarity arguments could also be posited, since the overlap of 
the $\pi$ systems of the two monomers would be expected to increase 
with planarity, and 11BP is more twisted (nearly perpendicular) 
than 22BP (37$^{\circ}$ dihedral angle\cite{fue16a}) and planar 22eBP. 
However, by linking pentacenes via an alkyne linker, a planar 6,6'-bipentacene 
has been synthesized\cite{leh10a} and shows no extra absorption 
in the visible (SI Fig. 1). Consequently, planarity is advantageous 
but not sufficient.
 
More sophisticated theories of chromophore interaction include Kasha's 
point-dipole model\cite{kas65a}, which can be used to explain the 
spectra of some acene derivatives\cite{coo17a,kum17a}. 
Kasha's model requires two bright excitations to interact, and can therefore 
explain the splitting of the high-energy absorption seen in the UV spectrum 
of 11BP, but not the emergence of a new absorption in a region of the 
spectrum where there was previously none. 
The new dimer absorption cannot result from the monomer 
$S_0\to S_1$ transition around 650nm splitting, since that would result in two 
absorption peaks with combined intensity equal to that of the monomer, whereas 
the dimer spectra of 22BP and 22eBP are more absorbent in the visible 
than the sum of their parts (see \figr{spec}).

However, intensity borrowing perturbation theory\cite{rob67a} is promising here,
as it describes how dark transitions can `borrow intensity' from bright 
states if they are coupled through a perturbation.
Since oscillator strength is conserved by the TRK sum 
rule\cite{tho25a,rei25a,kuh25a,atk11a}, the resulting new absorption in the 
spectrum is accompanied by a bright transition losing intensity. 
This is in keeping with our previous observation on the spectra of 22BP/22eBP 
where the appearance of a new visible absorption at 500nm is accompanied 
by a decrease in intensity of the UV peak. 
We also find that a previous computational study on 22BP which investigated 
its singlet fission properties and not the linear absorption spectrum, 
suggested the existence of a charge-transfer state around 440nm\cite{fue16a}. 
A reasonable hypothesis, is then, that a transition in the visible, 
possibly this CT excitation, is `borrowing intensity' from the UV in 
certain dimers like 22BP and 22eBP but not in 11BP.

\subsection{Intensity borrowing perturbation theory and PPP}
We develop a theory to describe the interactions between the monomer UV excitations and 
the low-energy CT excitation in a dimer by combining 
intensity borrowing perturbation theory\cite{rob67a} and PPP theory.
We also characterize the role that dimer bonding geometry
plays in modulating the strength of this interaction and, therefore, the appearance
(or not) of an additional peak in the absorption spectrum.

We first formally define our system in the language of perturbation theory. 
Similar to Kasha\cite{kas65a} and the ideas of symmetry-adapted perturbation 
theory\cite{jez94a}, for two monomers $n$ and $m$, the Hamiltonian is
\begin{align}
 \hat H = \hat H_n + \hat H_m +\hat V_{nm} \eql{hdef}
\end{align}
where $\hat H_n$ and $\hat H_m$ are the Hamiltonian operators of the monomers at infinite separation and $\hat V_{nm}=:\hat V$ is the intermonomer perturbation. 
Kasha approximated $\hat V$ as a dipole-dipole interaction and assumed 
all excitations were intra-monomer\cite{kas65a}, but here we make no 
\emph{a priori} assumptions about the nature of the excitations nor 
the functional form of $\hat V$. 
Instead, we choose to describe the overall system using PPP 
theory\cite{pop53a,pop55a,par53a,par56a} which we have already 
shown provides an accurate description of the electronic structure 
of these molecules. 

To define the monomer Hamiltonians within PPP 
theory,\cite{pop53a,pop55a,par53a,par56a,kou67a,ary13a} 
we let $\mathcal{M}$ be the set of atoms on monomer $m$ such that
\begin{align}
 \hat H_{m} = & \sum_{\mu \in \mM } \ep_\mu \hat n_{\mu} +  U_{\mu\mu} \hat n_{\mu,\uparrow} \hat n_{\mu,\downarrow} - \sum_{\langle \mu<\nu\in \mM\rangle} \sum_\sigma t_{\mu\nu} (\hat a^\dag_{\mu\sigma} \hat a_{\nu\sigma} + \hat a^\dag_{\nu\sigma} \hat a_{\mu\sigma}) \no\\
 & + \sum_{\mu<\nu \in \mM} \gamma_{\mu\nu} (\hat n_\mu - Z_{\mu})(\hat n_{\nu} - Z_{\nu}) \eql{hmon}
\end{align}
and likewise for monomer $n$, where $\hat n_{\mu}$ is the 
number operator for electrons on atom $\mu$,
\begin{equation}
\hat n_{\mu}=\sum_{\sigma=\{\uparrow,\downarrow\}}\hat n_{\mu,\sigma},
\end{equation}
where $\hat n_{\mu,\sigma}=\hat a_{\mu,\sigma}^{\dagger}\hat a_{\mu,\sigma}$,
and $\hat a_{\mu,\sigma}^{\dagger}$, $\hat a_{\mu,\sigma}$ are the creation 
and annihilation operators respectively for an electron of spin $\sigma$ 
on atom $\mu$. $Z_{\mu}$ is the local chemical potential, which is 1 for a carbon atom. 
$\epsilon_{\mu}$ is the on-site energy, which for a purely hydrocarbon 
chromophore we can set to zero without affecting the energies of excited states. 
$t_{\mu\nu}$ is the hopping term, where $\langle \rangle$ indicates the summation is 
over nearest neighbor pairs. 
$U_{\mu\mu}$ is the on-site (Hubbard) repulsion, and $\gamma_{\mu\nu}$ 
is the parameterized repulsion between an electron on atom $\mu$ 
and an electron on atom $\nu$ (for details see SI). 

Using the definition of the PPP Hamiltonian $\hat H$ for an arbitrary 
chromophore\cite{pop53a,pop55a,par53a,par56a,kou67a,ary13a} and 
\eqr{hmon}, we define the intermonomer perturbation as
\begin{align}
 \hat V & =:  \hat H - \hat H_n- \hat H_m \no \\
 & = -\sum_{\langle\mu \in \mM, \nu \in \mcN\rangle} \sum_{\sigma} t_{\mu\nu} (\hat a^\dag_{\mu\sigma} \hat a_{\nu\sigma} + \hat a^\dag_{\nu\sigma} \hat a_{\mu\sigma}) + \sum_{\mu \in \mM, \nu\in\mcN} \gamma_{\mu\nu} (\hat n_\mu - Z_{\mu})(\hat n_{\nu} - Z_{\nu}) \eql{vad}
\end{align}
Because the two chromophores are noninteracting at infinite separation, 
we can solve mean-field variants (Fock operators) of $\hat H_n$ and $\hat H_m$ separately for their corresponding 
molecular orbitals. 
The monomers are identical, 
allowing us to choose degenerate orbitals that are localized entirely on monomer $n$ or monomer $m$. 
These molecular orbitals are denoted $\phi_{ni}$ or $\phi_{mi}$ where the subscript indicates the $i^{th}$ MO
of monomer $n$ or $m$. 
In accordance with convention\cite{par56a} bonding orbitals (those occupied in the ground state) 
are numbered $1,2,3,\ldots$ from the HOMO downwards and antibonding orbitals $1',2',3',\ldots$ from the LUMO upwards. 
As noted by Pariser\cite{par56a}, care must be taken to be consistent with the relative signs of 
orbitals in order for the Coulson-Rushbrooke theorem\cite{cou40a} to be easily applied. 
For the perturbation analysis of the dimers considered here we choose every monomer orbital 
to have the same sign on the atom through which it is joined to the other monomer.

We are interested in linear absorption spectra and the dipole moment 
is a one-electron operator, so we consider only singly excited states 
in keeping with the original formulation of PPP. 
We denote single excitations $\ket{\Phi_{pi}^{qj'}}$ where $p$ and $q$ are 
either $n$ or $m$. We only consider singlet spin-adapted configurations\cite{sza89a} 
as triplet excitations are dark for hydrocarbons with minimal spin-orbit coupling. 
There are two types of single excitations. First, when $p=q$, $\ket{\Phi_{ni}^{nj'}}$ 
or $\ket{\Phi_{mi}^{mj'}}$, are intramolecular, local (Frenkel) excitations\cite{smi10a,bel13a,fue16a,li17a} 
that we denote as \cite{li17a} $\ket{\rLE_{ni}^{nj'}}=:\ket{\Phi_{ni}^{nj'}}$. 
For alternant hydrocarbons\cite{cou40a,pop53a,par56a} such as the molecules considered in this article, 
$\ket{\rLE_{ni}^{nj'}}$ and $\ket{\rLE_{nj}^{ni'}}$ are degenerate and we therefore define `plus' and `minus' excitations\cite{par56a} 
\begin{align}
 \ket{\rLE_{ninj}^{\pm}} = \frac{1}{\sqrt{2}} (\ket{\rLE_{ni}^{nj'}} \pm \ket{\rLE_{nj}^{ni'}})
\end{align}
where only `plus' excitations have nonzero transition dipole moment from the ground state.\cite{par56a}

The second type of excitation (when $p\ne q$), $\ket{\Phi_{ni}^{mj'}}$ or $\ket{\Phi_{mi}^{nj'}}$, are 
intermolecular, charge-transfer (CT) excitations\cite{bel13a,smi10a,li17a} which we denote 
$\ket{\rCT_{ni}^{mj'}}=:\ket{\Phi_{ni}^{mj'}}$. We appreciate that there are varying definitions 
and nomenclature for CT excitations in the literature\cite{smi10a,li17a,mul39b,fen13a,pie07a}, 
and here the definition corresponds to removing an electron from an orbital localized entirely 
on one monomer and placing it in an orbital localized on another monomer. 
Using the definition of the dipole moment in PPP theory\cite{par56a} and 
that orbitals on different monomers are spatially disjoint, we note that 
the CT excitations defined here are always dark, 
$\bra{\Phi_0}\hat \mu \ket{\rCT_{ni}^{mj'}} = 0$.

Our zeroth order eigenstates of interest are
thus the $\rLE$ and $\rCT$ states along with the ground state, 
$\ket{\Phi_0}$. 
The only bright states at zeroth order will be the PPP `plus' Frenkel excitations, $\ket{\rLE_{ninj}^{+}}$.

\subsection{First order perturbation}
Having obtained zeroth-order eigenstates of $\hat H_0$, we now consider how they are 
mixed by the perturbation $\hat V$ which occurs when the two monomer are covalently linked, 
and how this alters the linear absorption spectrum. 
Following standard perturbation theory\cite{atk11a}, we form the `good' degenerate 
eigenstates of $\hat H_0$ as linear combinations of excitations which we denote $A$ 
and $B$ in accordance with the irreducible representations (irreps) of the $C_2$ 
point group.\cite{atk11a} For Frenkel excitations we have
\bse
\begin{align} 
 \ket{\rLE_{ij}^{A,\pm}} = & \frac{1}{\sqrt{2}}(\ket{\Phi_{ninj}^\pm} + \ket{\Phi_{mimj}^\pm}), \\
 \ket{\rLE_{ij}^{B,\pm}} = & \frac{1}{\sqrt{2}}(\ket{\Phi_{ninj}^\pm} - \ket{\Phi_{mimj}^\pm}), 
\end{align}
\ese
and for CT
\bse
\begin{align} 
 \ket{\rCT_{ij}^{A}} = & \frac{1}{\sqrt{2}}(\ket{\Phi_{ni}^{mj'}} + \ket{\Phi_{mi}^{nj'}}), \\
 \ket{\rCT_{ij}^{B}} = & \frac{1}{\sqrt{2}}(\ket{\Phi_{ni}^{mj'}} - \ket{\Phi_{mi}^{nj'}}), 
\end{align}
\ese
which is similar to the linear combinations used in (for example) Kasha exciton theory\cite{kas65a}, that considers only the dipole-dipole interactions of Frenkel excitations. Note that in certain cases of degeneracy further linear combinations may be required but we do not find this necessary in what follows.

Since the perturbation $\hat V$ is symmetric under all the symmetry operations 
of the dimer, only like irreps can mix. We are therefore interested in the UV 
Frenkel excitations $\ket{\rLE_{14}^{A,+}}$ and $\ket{\rLE_{14}^{B,+}}$ mixing 
with the dark charge-transfer excitations $\ket{\rCT_{11}^{A}}$ and $\ket{\rCT_{11}^{B}}$ 
respectively. Using standard electronic structure theory algebra\cite{sza89a} we find
\begin{align}
\bra{\rLE_{14}^{A,+}} \hat V \ket{\rCT_{11}^{A}} = & 0 \eql{fbctba} \\
\bra{\rLE_{14}^{B,+}} \hat V \ket{\rCT_{11}^{B}} = & \frac{1}{2\sqrt{2}}\left( t_{n4'm1'} + t_{n4m1} + t_{m4'n1'} + t_{m4n1} \right)
\eql{fbctb}
\end{align}
where $t_{\mathrm{nimj}}$ is an element of the one-electron, hopping matrix in the 
molecular orbital basis. 
Note that due to the NDO approximation present within PPP theory, there are no two-electron terms in \eqr{fbctb}. 
Equation \ref{eq:fbctba} shows that $\ket{\rCT_{11}^A}$ cannot borrow intensity from 
$\ket{\rLE_{14}^{A,+}}$ (to first order), and we therefore need only consider the 
corresponding $B$ excitations. This result is particularly convenient for planar or 
near-planar structures such as 22BP and 22eBP where dipole moment arguments show that 
only excitations of $B$ symmetry are likely to have significant oscillator strength.

To quantify the extent of mixing, we further simplify the right hand side of \eqr{fbctb}. 
Using the Coulson-Rushbrooke theorem\cite{cou40a} and the definition of the sign of monomer 
orbitals (see above) we find that $t_{n4'm1'} =  t_{n4m1}$ and $t_{m4'n1'} = t_{m4n1}$. 
We also find from the $C_2$ symmetry of the dimer that $t_{n4'm1'} = t_{m4'n1'}$ and $t_{n4m1} = t_{m4n1}$. 
We can then evaluate the one unique matrix element to obtain
\begin{align}
\bra{\rLE_{14}^{B,+}} \hat V \ket{\rCT_{11}^{B}} = \sqrt{2} t_{\nu^*\sigma^*} c_{n1,\nu^*} c_{m4,\sigma^*} \eql{rtbeta}
\end{align}
where $c_{ni}$ represent the expansion coefficients for the monomer 
orbitals in the atomic orbital basis, $\phi_{ni} = \sum_\lambda c_{ni,\lambda} \chi_\lambda$. 
Due to the nearest-neighbor nature of $t$, the only relevant expansion coefficients 
in \eqr{rtbeta} become those associated with the dimer bonding position, 
i.e., $\nu^*$ and $\sigma^*$. 
We further note that $t_{\nu^*\sigma^*}$ will in general be proportional to 
$\cos(\theta)$ where $\theta$ is the dihedral angle between the two monomers.
Finally, using intensity borrowing perturbation theory\cite{rob67a}, 
we find that the dipole moment of the CT excitation is, at first order,
\begin{align}
\bra{\Phi_0} \hat \mu \ket{\rCT_{11}^{B,(1)}} \simeq -\bra{\Phi_0} \hat \mu \ket{\rLE_{14}^{+,B}} \frac{\sqrt{2} t_{\nu^*\sigma^*} c_{n1,\nu^*} c_{m4,\sigma^*}}{E(\rCT_{11}^B) - E(\rLE_{14}^{+,B})} 
\eql{main}.
\end{align}
and the intensity of the new absorption will be proportional to the square of the dipole moment\cite{atk11a}.

\section{Application to bipentacenes}
\label{sec:molorb}

Having arrived at an expression to estimate intensity borrowing to first order in \eqr{main}, we undertake
an investigation of the extent to which this phenomenon is observed in bipentacenes.

For 11BP a monomer calculation gives $c_{m4,\sigma^*=1}=0$, such that 
$\bra{\Phi_0} \hat \mu \ket{\rCT_{11}^{B,(1)}}$ and there is no new 
low-energy absorption. For 22BP a monomer calculation gives $c_{m4,\sigma^*=2}\neq0$, and we find
\begin{align}
 -\sqrt{2} t_0 \cos(\theta) c_{n1,\nu^* = 2} c_{m4, \sigma^* =2} = & -0.0860 \mathrm{\ eV} \\
 E(\rCT_{11}^B) = & 3.58 \mathrm{\ eV} \\
 E(\rLE_{14}^{B,+}) = & 3.86 \mathrm{\ eV}
\end{align}
where we use the atom numbering in \figr{struc} and set $t_0 = 2.2$~eV and $\theta = 37^{\circ}$. Consequently, 
\begin{align}
\frac{\bra{\rLE_{14}^{B,+}} \hat V \ket{\rCT_{11}^{B}}}{E(\rCT_{11}^B) - E(\rLE_{14}^{+,B}) } = 0.30.
\end{align}
The perturbation therefore corresponds to $0.30^2 \simeq 9\%$ of the UV peak intensity being borrowed.
We emphasize that the 9\% intensity borrowing calculated here is 
qualitative as we are neglecting all second-order contributions arising
from mixing with other states. 
We further note that although the perturbation is sufficiently significant, leading to 
visible change in color, is still weak as evidenced by the UV absorption in 22BP
and 22eBP continuing to be the dominant absorption.
We further calculate the energy of the low-energy absorption from the first 
order correction to the energy of the CT state, which is 
$\bra{\rCT_{11}^B} \hat V \ket{\rCT_{11}^B} = -(n1n1|m1'm1') = -0.91 $eV, corresponding 
to the Coulomb attraction of an electron in the LUMO of one molecule and 
a hole in the HOMO of the other. 
This gives $E_{\rCT_{11}^B}^{(1)} = 2.67$ eV (464 nm), close to the experimentally 
observed new transition at 496 nm.

The dihedral angle arguments advanced earlier for the strength of the absorption 
also explain the more intense absorption seen in 22eBP. Treating the alkene 
linker to be a molecular `wire' through which the monomers interact, we estimate
the new absorption in 22eBP to be $\cos^2(0^\circ)/\cos^2(37^\circ) \simeq 1.57$ 
times greater than the absorption in 22BP. This is experimentally verified by 
the stronger visible absorption peak that appears in the spectra of \figr{spec}A.

We find that the conclusions reached through intensity borrowing and 
perturbation theory (\eqr{main}) above can be anticipated by examining 
the nodal structure of the relevant monomer orbitals. 
In \figr{horbs}A, we present the top four HOMOs of TIPS-Pentacene 
obtained from an RHF calculation that are qualitatively similar 
to the orbitals from a PPP calculation. The HOMO--3 orbital 
has nodes on every long-axis carbon (red arrows in \figr{horbs}A), 
as is the case for all acenes\cite{par56a}, and has been observed 
experimentally.\cite{soe09a} 
The HOMO, HOMO--1 and HOMO--2 all have nodes in the horizontal ($xz$) 
plane but the HOMO--3 does not.
Now, consider joining together two monomers by the $1,1'$ and $2,2'$ positions, 
shown schematically in \figr{horbs}B. 
For 11BP in \figr{horbs}B(i), we see that the HOMO has nonzero amplitude 
at the $1$ position, but the HOMO--3 has zero amplitude such that 
the perturbation $c_{n1,\nu^*}c_{m4,\sigma^*}=0$, leading to no 
new low-energy absorption peak in the spectrum. 
For 22BP, as shown in \figr{horbs}B(ii), both the HOMO and HOMO--3 
have substantial amplitude, $c_{n1,\nu^*}c_{m4,\sigma^*}\neq 0$ 
and a new low-energy absorption peak is predicted, and verified both 
by the experimental spectrum and from PPP calculations (\figr{spec}). 

Using similar arguments, once can show that indirect mixing 
(via another excitation) is likely to be minimal unless the monomers 
are joined via a short-axis carbon. 
The same inferences can be drawn by examining the monomer LUMOs using
the Coulson-Rushbrooke theorem\cite{cou40a}, as shown in the SI.

\begin{figure}[tb]
 \includegraphics[width=\textwidth]{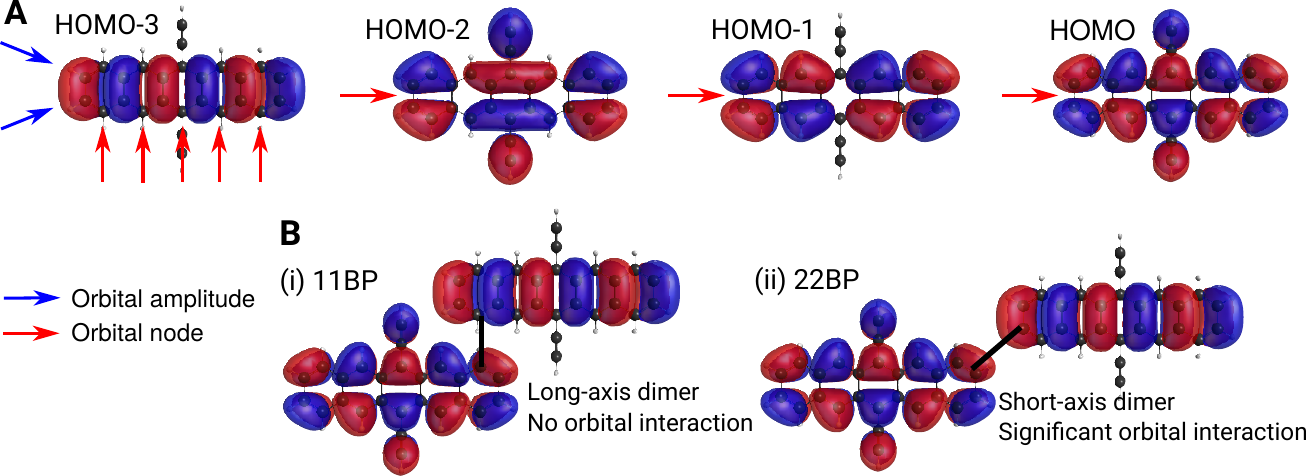}
 \caption{\textbf{A} Top four HOMOs of TIPS-pentacene. Transitions from the HOMO--3 contribute to the intense UV absorption in the monomer and those from the HOMO to a low-energy CT state. \textbf{B} shows how examining nodal structure of orbitals can explain the emergence of an extra absorption. Bonding along a long-axis carbon as for 11BP in \textbf{B}(i) leads to no significant interaction between the HOMO--3 on one monomer and the HOMO on the other. Conversely, bonding via a short-axis carbon as for 22BP in \textbf{B}(ii) leads to a significant orbital interaction. For clarity the dimers are drawn planar and with an elongated inter-monomer bond (thick black line) and long/short axes are defined in \figr{struc}\textbf{A}.}
 \figl{horbs}
\end{figure}

The intensity borrowing and molecular orbital arguments above show that 
acene monomers must be joined via short-axis carbons in order to 
observe a new, low-energy absorption. 
Since this analysis can, in theory, be performed for arbitrarily 
many monomers, we would expect this result to hold for oligomers 
and polymers as well as dimers. In addition, the MO arguments suggest 
that joining acenes via a long-axis carbon or \emph{both} short-axis carbons 
will \emph{not} lead to enhanced low-energy absorption, since the HOMO has 
a long-axis ($xz$) nodal plane whereas the HOMO--3 does not,
leading to no constructive interaction between the relevant orbitals.

We are now ready to construct a simple design rule to make 
acene dimers, oligomers and polymers that will exhibit a new 
low-energy (visible) absorption peak in their spectra:
\begin{quote}
    Join the monomers via a short-axis carbon, and 
    avoid a long-axis symmetry plane passing through 
    adjacent monomers.
\end{quote}
This is the main result of the article and is summarized in \figr{rule}. 
\begin{figure}[h]
 \includegraphics[width=0.8\textwidth]{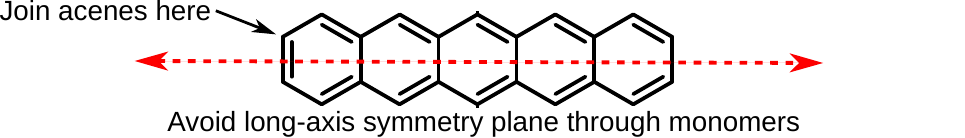}
 \caption{Design rule identifying bonding motifs in 
 acene-based dimers and oligomers that exhibit an additional low-energy
 absorption}
 \figl{rule}
\end{figure}

Clearly, for the low-energy absorption to be significant there 
must be significant interactions between the monomers' $\pi$ systems -- 
they must be directly bonded, otherwise held very close together, 
or be connected via a conjugated linker (such as in 22eBP). 
Furthermore, if the design rule is satisfied, 
increasing planarity will increase the intensity of the extra absorption. 

\section{Application to general acenes}
\label{sec:app}
We have already demonstrated that the design rule proposed in the previous 
section can explain the presence or absence of a visible absorption peak in
the spectra of 11BP, 22BP, and 22eBP.
Here, we demonstrate the broad applicability of this rule by anticipating the 
experimental UV-Vis spectra of acene dimers, trimers, oligomers, and polymers, 
as well as a variety of hetero-atom substituted derivatives made for a broad 
range of applications including as a demonstration of 
organic synthesis \cite{aki60a,yam16a,rud14a,bul13a}, 
for organic semiconductors \cite{gey15a,bal11a,leh10a}, liquid crystals \cite{lon01a}, 
polymer synthesis \cite{bho13a}, sensors \cite{gu15a}, photovoltaic 
applications \cite{bar11a,kum16a,coo16a}, and one to explore structure-spectrum 
relationships \cite{fri48a}. 

\begin{figure}[tb!]
\includegraphics[width=\textwidth]{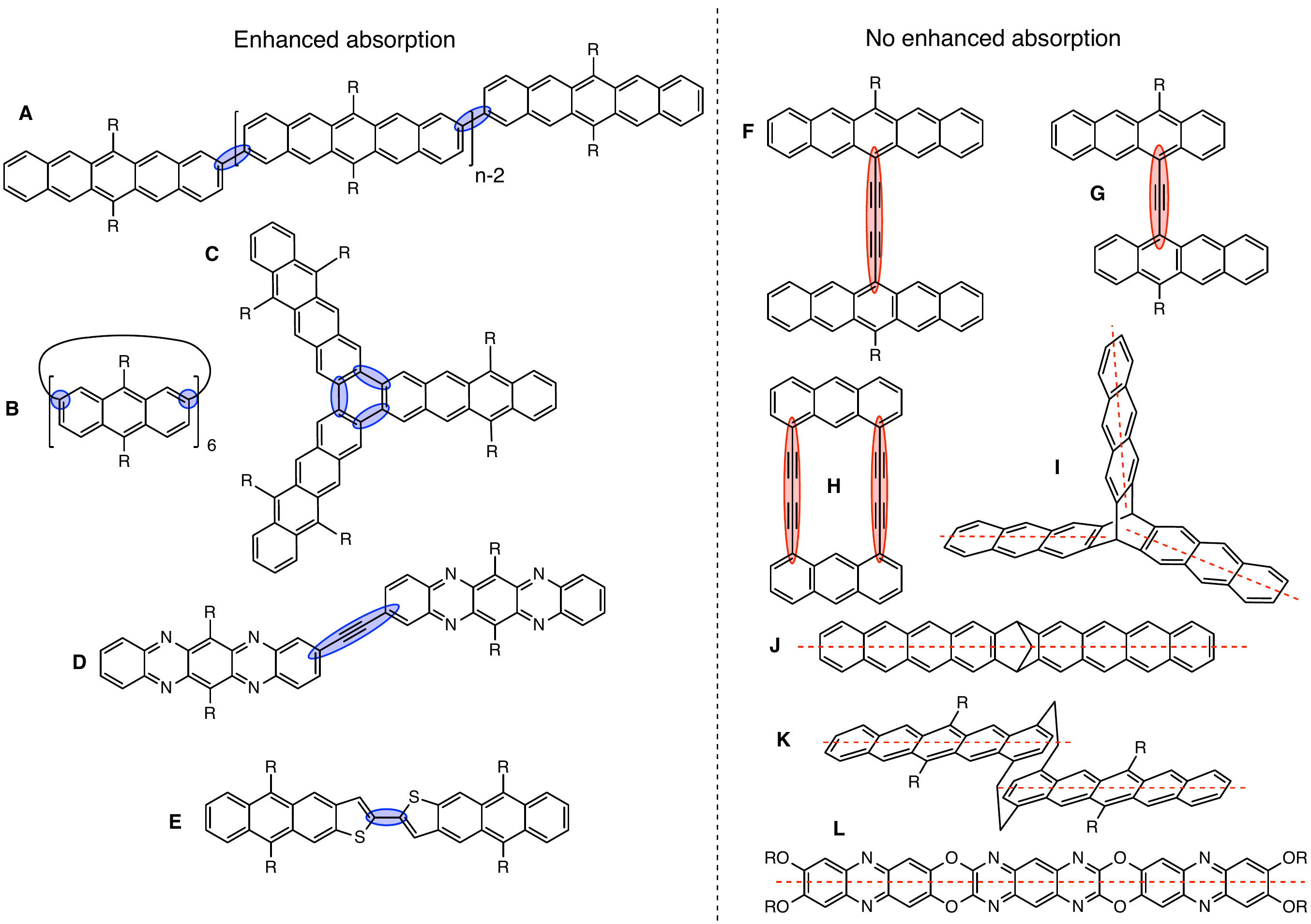}
\caption{Illustrative examples of acene derivatives, showing how the design rule correctly predicts 
    the presence or absence of an extra, low-energy absorption. Left: molecules bonded via a 
short-axis carbon (blue oval) and without a horizontal ($xz$) symmetry plane through adjacent 
monomers all show enhanced low-energy absorption in their experimental UV-vis spectra. 
Right: molecules bonded via a long-axis carbon (red oval), or such that the horizontal 
symmetry plane is preserved through adjacent monomers (red dashes), none of which show 
enhanced absorption compared to their corresponding monomer(s). 
The design principle correctly applies to acene dimers, oligomers and polymers such as 
A \cite{kum16a}, B \cite{yam16a}, F \cite{leh10a}, G \cite{bar11a} and H \cite{aki60a}, 
complex geometries such as C \cite{rud14a}, I \cite{lon01a,bho13a}, J\cite{coo16a} and 
K \cite{bul13a}, and to heteroatom-substituted dimers D \cite{gey15a}, E \cite{bal11a}, 
and L \cite{gu15a}. R denotes a solubilising or stabilising group.}
\figl{applic}
\end{figure}

A small selection of these are drawn in \figr{applic} grouped according to 
whether or not their spectra have enhanced low-energy absorption compared 
to their corresponding monomers. 
Our design rule correctly predicts the presence or absence of an 
extra absorption in all cases, including dimers, oligomers and 
polymers of naphthalene~\cite{fri48a}, anthracene (\figr{applic}B 
and \figr{applic}H~\cite{yam16a,aki60a}), tetracene (\figr{applic}G \cite{bar11a})  
and pentacene (Fig.~1, \figr{applic}A and 
\figr{applic}F \cite{kum16a,san15a,fue16a,leh10a}), 
thereby explaining the structure-spectrum phenomenon first observed in 1948 \cite{fri48a}. 
It also correctly predicts the presence or absence of an extra absorption for 
unusual and complex geometries such as starphenes (\figr{applic}C \cite{rud14a}), 
iptycenes (\figr{applic}I \cite{lon01a,bho13a}), bridged 
dimers (\figr{applic}J \cite{coo16a}), and cyclophanes (\figr{applic}K \cite{bul13a}).
Further, we find this design rule also holds in heteroatom-substituted 
acene derivatives such as the tetra-aza-pentacene dimer in \figr{applic}D\cite{gey15a}, 
the anthrathiophene dimer in \figr{applic}E\cite{bal11a} and the aza-anthracene 
trimer \figr{applic}L\cite{gu15a} 
(compared to its corresponding monomers\cite{lee10a,gaw11a}). 
This unexpected applicability of the design rule can be attributed to 
the `striped' orbitals in \figr{horbs}A that will be solutions to the 
H\"uckel Hamiltonian of the heteroatom-substituted acene in cases where 
only the diagonal energies ($\ep_\mu$) are perturbed and not the 
off-diagonal ($t_{\mu\nu}$) terms. In general, this will hold when the 
the heterosubstitution is only on long-axis carbons. 
For the thiophene derivative, standard organic chemistry\cite{cla01a} 
suggests that the S heteroatom is of a similar size to a C=C double bond, 
making thiophene qualitatively similar to benzene. 
With this reasoning the anthrathiophene dimer can be 
likened to a $2,2'$-tetracene dimer, explaining the new low-energy absorption. 

Finally, we note that it is challenging to apply the design rule described
here to oligomers where two or more moieties overlap in 
such a way that through-space interaction become important, 
such as the cross-conjugated dimers in Refs~\citenum{zir15a} and \citenum{zir16a}. 
In these cases a more sophisticated formulation of the hopping term 
than nearest-neighbor interactions will be required for 
PPP theory (and the intensity borrowing arguments) to be applied.

\section{Conclusions}
\label{sec:conc}
This article uses PPP theory and intensity borrowing perturbation theory to 
construct a simple rule for predicting and explaining the low-energy 
absorption spectra of acene derivatives. The resulting testable and 
experimentally verifiable design rule has been found to hold in a 
large variety of dimers, trimers, oliogmers and polymers, 
including those with heteroatom substitution and unusual bonding geometries.
This is particularly useful because it allows for the presence 
or absence of a new low-energy absorption to be determined from 
the \emph{monomer} orbitals alone, without having to simulate 
the electronic structure of each dimer or oligomer separately.
This \emph{a priori} design of highly absorbent molecules, of 
which this is a proof-of-concept, has significant implications 
for photovoltaic design, where some organic solar photovoltaics 
are constrained by the small diffusion length of excitons compared 
to the thickness of material required to absorb a significant 
fraction of visible light\cite{for05a}. Developing more 
absorbent molecules that retain the necessary photophysical 
charge-transport properties could allow for thinner, flexible, 
cheaper and more efficient solar cells.

\begin{acknowledgement}
We thank Roald Hoffmann and Tao Zeng for helpful discussions and acknowledge spectra from Dan Lehnherr and Rik Tykwinski. We also thank Robert A.\ DiStasio Jr.\ and Stuart Althorpe for helpful comments on the manuscript. We thank the Nuckolls lab for use of their UV-vis spectrophotometer. NA acknowledges funding from NSF CAREER grant (Award No. CHE-1555205) and a Sloan Foundation Research Fellowship. TJHH acknowledges funding from NSF EAGER grant (Award No. CHE-1546607) and from Jesus College, Cambridge. This work used the Extreme Science and Engineering Discovery Environment (XSEDE), supported by National Science Foundation grant number ACI-1053575. L.M.C. acknowledges support from the Office of Naval Research Young Investigator Program (Award N00014-15-1-2532), ACS Petroleum Research Fund, 3M Non-Tenured Faculty Award, Arthur C. Cope Scholar Award, and Cottrell Scholar Award. SNS thanks the NSF for GRFP (DGE 11-44155). This research used resources of the Center for Functional Nanomaterials, which is a U.S.\ DOE Office of Science Facility, at Brookhaven National Laboratory under Contract No. DE-SC0012704. 
\end{acknowledgement}


\bibliography{refbig}

\end{document}


\bibliographystyle{achemso}

\title{Supporting information for: \\ Anticipating acene-based chromophore spectra with molecular orbital arguments}
\author{Timothy J.\ H.\ Hele}
\email{tjhh2@cam.ac.uk}
\affiliation{Cavendish Laboratory, JJ Thomson Avenue, Cambridge University, CB3 0HE, UK}
\author{Eric G. Fuemmeler}
\affiliation{Baker Laboratory, 259 East Avenue, Cornell University, Ithaca, NY 14850, USA}
\author{Samuel N. Sanders}
\author{Elango Kumarasamy}
\affiliation{Department of Chemistry, Columbia University, New York, NY 10027, USA}
\author{Matthew Y. Sfeir}
\affiliation{Center for Functional Nanomaterials, Brookhaven National Laboratory, Upton, NY 11973, USA}
\author{Luis M. Campos}
\email{lcampos@columbia.edu}
\affiliation{Department of Chemistry, Columbia University, New York, NY 10027, USA}
\author{Nandini Ananth}
\email{ananth@cornell.edu}
\affiliation{Baker Laboratory, 259 East Avenue, Cornell University, Ithaca, NY 14850, USA}
\date{\today}

\maketitle

\tableofcontents

\section{Computational details}
All calculations were performed \emph{in vacuo} as the ground states of alternant hydrocarbons such as pentacene have no dipole (or multipole) and are therefore not expected to show a solvatochromic shift.

Our first task is to compute the ground-state structures of these molecules. Consistent with previous calculations on TIPS-pentacene derivatives, we convert the Si($i$-Pr)$_3$ groups to hydrogens as they do not significantly affect the electronic structure of the chromophore \cite{fue16a}. Similar to previous calculations on the pentacene monomer (without the TIPS groups) \cite{zen14a}, we optimize the monomer structure using complete active-space self-consistent field (CASSCF), here at the 8o8e level with orbitals chosen by energy (four highest HOMOs and four lowest LUMOs), using a restricted Hartree-Fock (RHF) reference and no state-averaging. As in previous work \cite{zen14a} we used a trimmed cc-pVTZ basis, with the $f$ functions removed from carbon and the $d$ functions from hydrogen. The 22BP structure was taken from Ref.~\citenum{fue16a}. Due to computational cost, 11BP, 22eBP, and 66yBP were optimized using DFT with the B3LYP functional in a 6-31G$^*$ basis. Coordinates are given in section~\ref{sec:coord} of the SI.

High-level monomer calculations used CASSCF and generalized multiconfiguational quasidegenerate perturbation theory (GMC-QDPT) with the same basis as for optimization. CASSCF calculations were state-averaged over the lowest three states of a given symmetry, and GMC-QDPT calculations used the standard intruder state avoidance parameter of 0.02$E_h^2$ and Granovsky's zeroth-order Hamiltonian for XMC-QDPT \cite{gra11a}. Pictures of monomer MOs were generated using the same basis from a restricted Hartree-Fock (RHF) calculation, have the same qualitative structure as from a PPP calculation, and are ordered according to their PPP energy. 

Similar to previous research \cite{ary10a} for PPP calculations the hopping parameter $t_{\mu\nu}$ was set to $2.2$eV for single bonds, $2.4$ eV for aromatic bonds, and $2.8$ eV for triple bonds, scaled by a cosine of the dihedral angle for inter-monomer bonds \cite{ven06a}. The electronic repulsion terms were parameterized using the Mataga-Nishimoto form \cite{mat57a}: $V_{\mu\nu}=U/(1+\frac{R_{\mu\nu}}{r_0})$, where the Hubbard parameter $U= 8$ eV \cite{ary10a} and $r_0 = 1.634$\AA, which was found to give good agreement for the pentacene monomer (see Table~\ref{tab:spec} below). 
For highly non-planar molecules, other NDO methods such as INDO could be used for spectral prediction \cite{zol03a,ges03a} but conventional PPP theory is found to be sufficiently accurate for the molecules studied (see Fig.~2 of the main article).
For plotting the results of PPP calculations, the excited state energies were not shifted to agree with experiment. Peaks were broadened used a full width at half maximum value of 20 nm, and do not include vibrational progressions.  PPP calculations were performed using in-house code finding the lowest 25 singly excited states.  All other electronic structure calculations were performed using the GAMESS-US package \cite{sch93a,gor05a}, using the standard input parameters unless otherwise stated, and used XSEDE resources \cite{tow14a}. Orbitals are plotted using Macmolplt and a contour value of 0.01.

In Table~\ref{tab:spec} we compare the results of GMC-QDPT, PPP theory and experiment for relevant low-lying excited states.

\begin{table}[tb]
\centering
\resizebox{\textwidth}{!}{
\begin{tabular}{c|c|c|c|c|c|c}
Exp. abs. & GMC-QDPT & GMC-QDPT & GMC-QDPT & PPP state & PPP excitations & PPP \\
$\lambda$ (nm) & state & excitations & $\lambda$ (nm) & &  & $\lambda$ (nm) \\ \hline
642 & $1B_{2u}$ & $0.94\ket{\Phi_1^{1'}}$ & 597 &$1B_{2u}$ & $ 0.99\ket{\Phi_1^{1'}}$ & 662 \\
440 & $1B_{3u}$ & $0.72\ket{\Phi_4^{1'}} - 0.58\ket{\Phi_1^{4'}}$ & 460 &$1B_{3u}$& $ 0.67(\ket{\Phi_4^{1'}} - \ket{\Phi_1^{4'}})$ & 385 \\ 
310 & $3B_{3u}$ & $0.59\ket{\Phi_4^{1'}}+ 0.71\ket{\Phi_1^{4'}}$& 307 &$2B_{3u}$& $0.69(\ket{\Phi_4^{1'}}+ \ket{\Phi_1^{4'}})$ & 317
\end{tabular}
}
\caption{Energies of selected excited states of TIPS-pentacene computed using 8o8e CASSCF with GMC-QDPT, and PPP theory (see computational details). PPP orbital numbering is used \cite{par56a}, where $\ket{\Phi_i^{j'}}$ is a singlet spin-adapted excitation from monomer orbital $i$ to monomer orbital $j'$. For comparison, orbitals are ordered according to their PPP energy, and their signs are chosen to be consistent with the Coulson-Rushbrooke theorem \cite{cou40a,par56a}. The GMC-QDPT $2B_{3u}$ state is predominantly composed of double excitations and is therefore not captured by PPP theory.}
\label{tab:spec}
\end{table}

Examining the high-level and PPP calculations in Table~\ref{tab:spec}, we find that PPP is more accurate than GMC-QDPT for the lowest-energy transition, less accurate for the dark $1B_{3u}$ state and both methods capture the UV excitation around 310nm to within 10nm. Importantly, the excitations corresponding to the excited states in Table~\ref{tab:spec} are the same in both GMC-QDPT and PPP and weighted by similar coefficients, showing that despite the simplicity of PPP theory, it provides an accurate description of these excited states.

\clearpage
\section{Supplementary Figures}
\setcounter{figure}{0}

\begin{figure}[h]
\centering
\includegraphics[width=120mm]{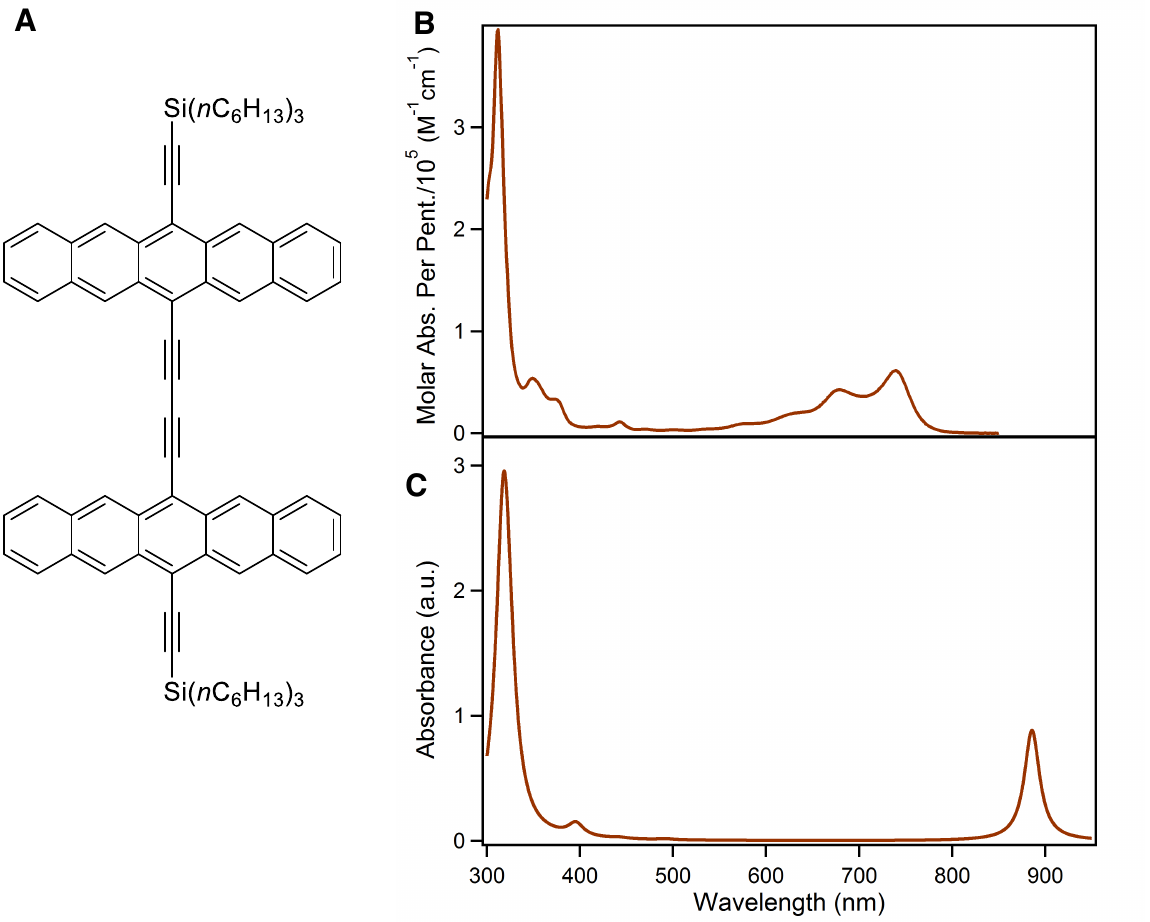}
\caption{Structure and spectra of the planar bipentacene 66yBP. \textbf{(A)}, Chemical structure of the alkyne-linked $6,6'$-bipentacene \cite{leh10a}. Neither the experimental spectrum \cite{leh10a} (\textbf{B}) nor computed spectrum using PPP theory (\textbf{C}) show an extra low-energy absorption compared to TIPS-P. This is despite the ability of the dimer to lie flat, and the large inter-chromophore interaction which causes the lowest-energy excitation to be redshifted by approximately 100nm compared to monomeric TIPS-Pentacene.}
\figl{66yBP}
\end{figure}

\begin{figure}[h]
\centering
\includegraphics[width=150mm]{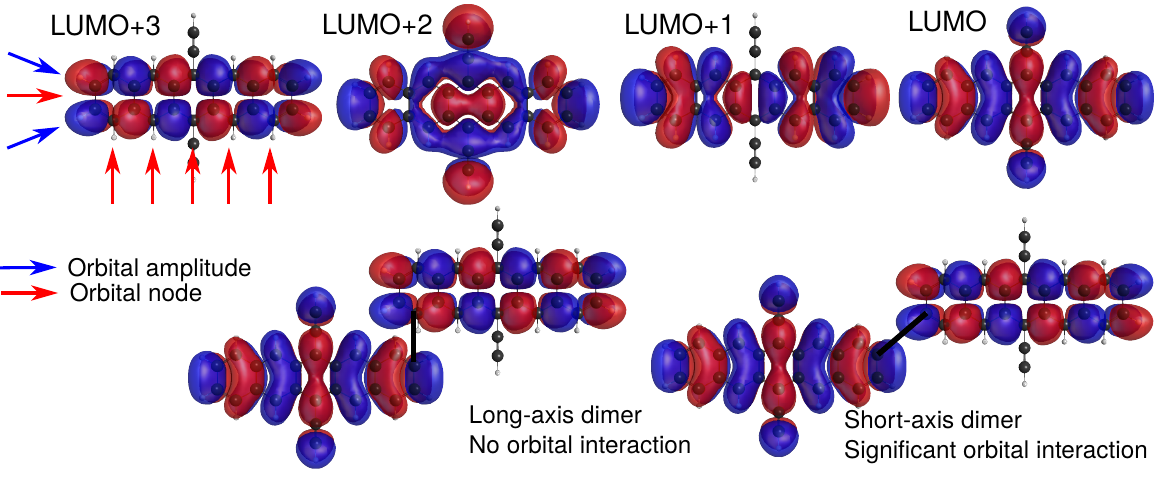}
\caption{\textbf{Monomer LUMOs.} LUMOs of TIPS-pentacene, showing how the LUMO+3, like the HOMO--3, has nodes on all long-axis tertiary carbons. Furthermore, the LUMO+3 has a node in the long-axis ($xz$) symmetry plane, whereas none of the other LUMOs shown do.}

\end{figure}

\clearpage

\section{Experimental methods}

\subsection{General Methods} 
All commercially obtained reagents/solvents were
used as received; chemicals were purchased from Alfa
Aesar$^{\circledR}$, Sigma-Aldrich$^{\circledR}$, Acros
organics$^{\circledR}$, TCI America$^{\circledR}$,
Mallinckrodt$^{\circledR}$, and Oakwood$^{\circledR}$
Products, and were used as received without further purification. Unless
stated otherwise, reactions were conducted in oven-dried glassware under
argon atmosphere. \textsuperscript{1}H-NMR and \textsuperscript{13}C-NMR
spectra were recorded on Bruker 400 MHz (100 MHz for
\textsuperscript{13}C) and on 500 MHz (125 MHz for
\textsuperscript{13}C) spectrometers. Data from the
\textsuperscript{1}H-NMR and \textsuperscript{13}C spectroscopy are
reported as chemical shift ($\delta$ ppm) with the corresponding integration
values. Coupling constants (\emph{J}) are reported in hertz (Hz).
Standard abbreviations indicating multiplicity were used as follows: s
(singlet), b (broad), d (doublet), t (triplet), q (quartet), m
(multiplet) and virt (virtual). The mass spectral data for the compounds
were obtained from XEVO G2-XS Waters$^{\circledR}$ equipped with a
QTOF detector with multiple inlet and ionization capabilities including
electrospray ionization (ESI), atmospheric pressure chemical ionization
(APCI), and atmospheric solids analysis probe (ASAP). The base peaks
were usually obtained as {[}M{]}\textsuperscript{+} or
{[}M+H{]}\textsuperscript{+} ions.

Anhydrous solvents were obtained from a Schlenk manifold with
purification columns packed with activated alumina and supported copper
catalyst (Glass Contour, Irvine, CA). All reactions were carried out
under argon unless otherwise noted.

\subsection{General protocol for the synthesis of bipentacenes}
The general synthetic scheme is similar to the previously-reported synthesis of 22BP\cite{kum16a,san15a,fue16a}.
\begin{figure}[h]
\includegraphics[width=\textwidth]{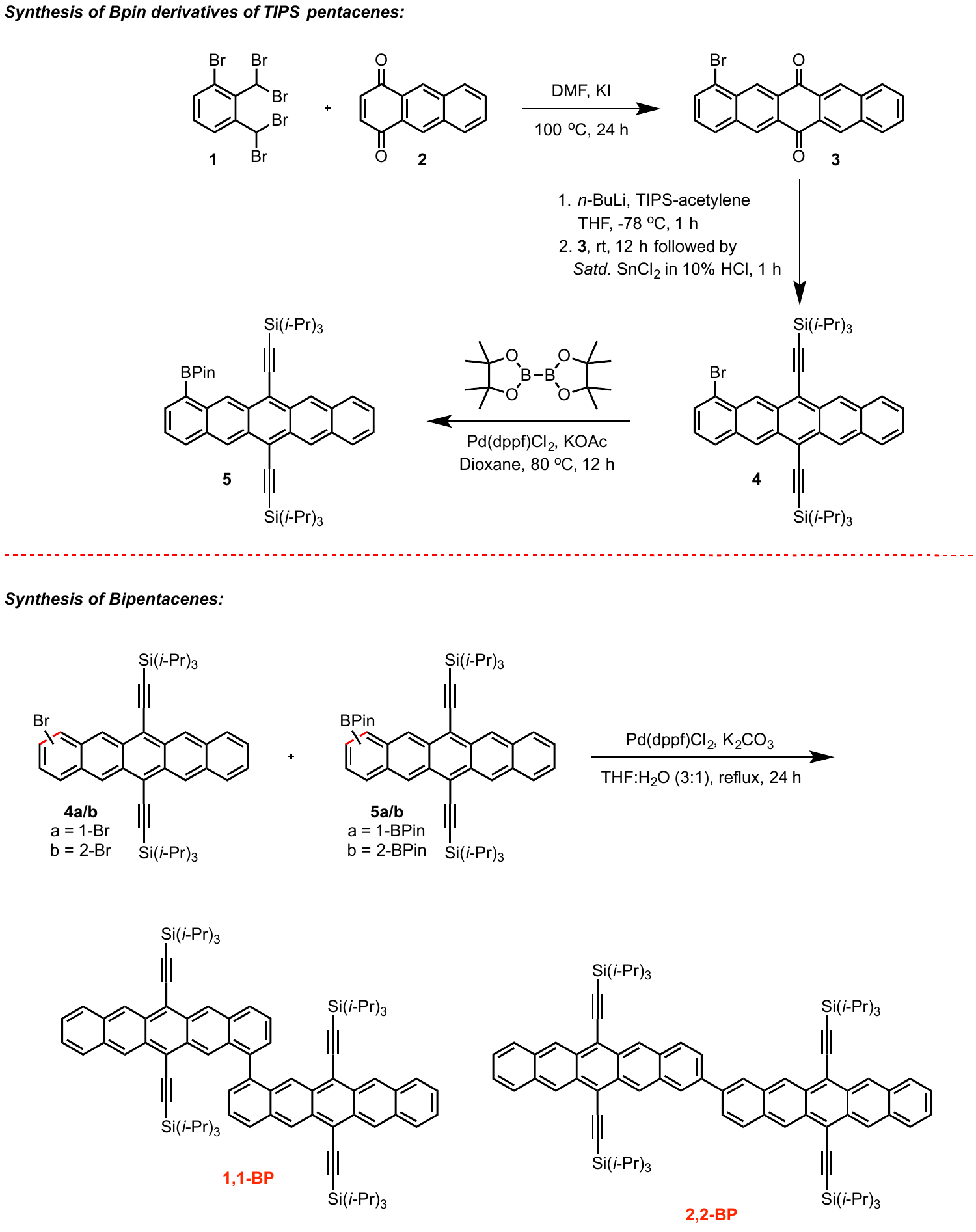}
\end{figure}
For more details on the synthesis of 2,2-BP and other intermediates please refer to our previous publications \cite{kum16a,san15a,fue16a}.
\clearpage

\subsection{General protocol for the synthesis of trans-ethylene-2,2-bipentacene}
\begin{figure}[h]
\includegraphics[width=\textwidth]{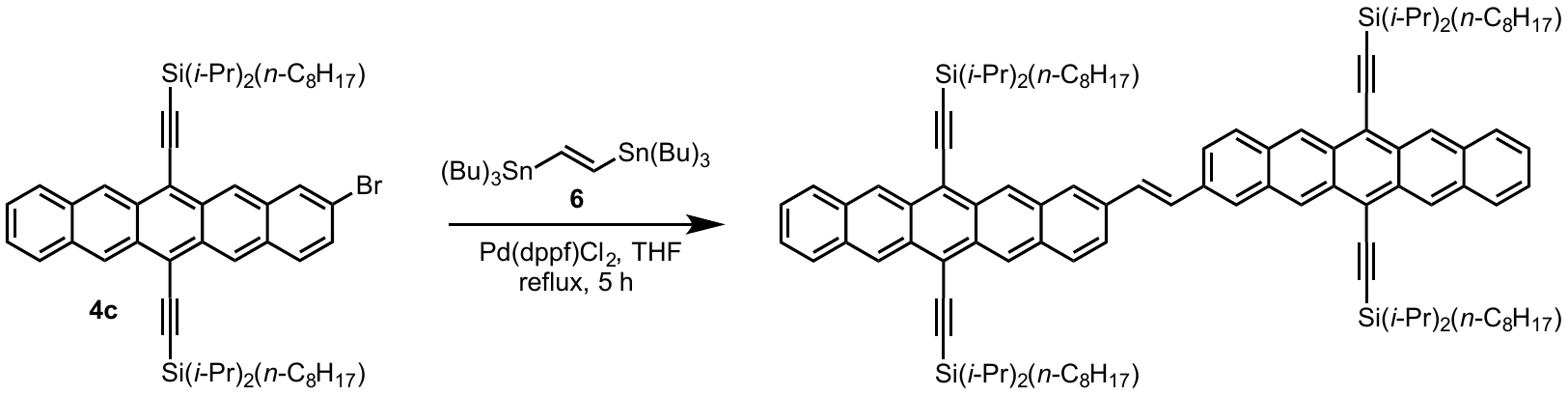}
\end{figure}

\subsection{General procedure for the synthesis of BPin derivatives of pentacenes}
\subsubsection{Molecule 3}

\begin{figure}[h]
\includegraphics[width=.3\textwidth]{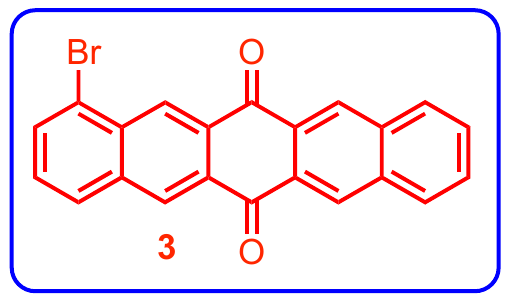}
\end{figure}
A mixture of 1,2-bis(dibromomethyl)benzene derivative \textbf{1} (1.1
\emph{equiv}) and anthracene-1,4-dione (1.0 \emph{equiv.}) and KI (4.0
\emph{equiv.}) in dry DMF was heated to 100$^\circ$ C and maintained for 24 h.
After the reaction, the mixture was cooled to room temperature and
diluted with methanol. The slurry was filtered and the solid was washed
with water and methanol. The resulting 6,13-pentacenequinone was
collected and dried. The crude product was directly used for next step
without further purification.

Due to limited solubility of the 6,13-pentacenequinones no
characterization was carried out.

\subsubsection{Molecule 4}
\begin{figure}[h]
\includegraphics[width=.3\textwidth]{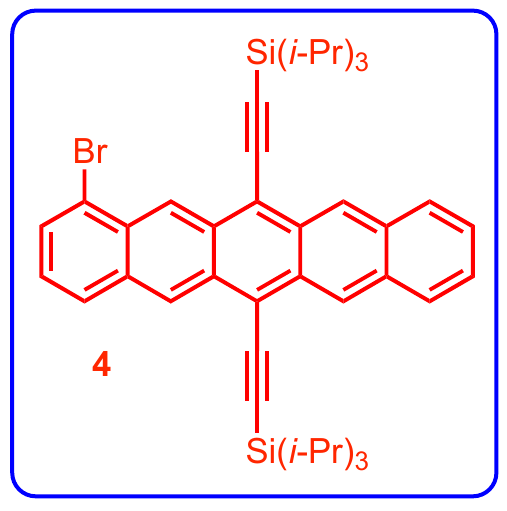}
\end{figure}
To a solution of triisopropylsilylacetylene (3.5 \emph{equiv.}) in dry and
degassed THF (25 mL) in 200 mL Schlenk flask at -78 $^\circ$C added
\emph{n}-butyl lithium (3.4 \emph{equiv.}, 2.5 M in hexanes). This
solution was allowed to stir at -78 $^\circ$C for 1 h followed by the addition
of \textbf{3} (4.0 g, 1.0 \emph{equiv.}) under positive argon flow. The
solution was allowed to warm to rt and stirred overnight (16 h) or until
solid pentacenoquinone was no longer observed. To this clear, deep
yellow solution was added of a saturated solution of tin (II) chloride
dihydrate in 10\% aqueous HCl solution (50 mL) during which the solution
turned deep blue. The resulting mixture was stirred at rt for 1 h under
dark and filtered over a pad of silica. The solid was washed with DCM
and the combined organic layer was washed with water (2 $\times$— 200 mL), dried
over \emph{anhyd.} Na\textsubscript{2}SO\textsubscript{4}, filtered and
the solvent was removed under reduced pressure to get the crude product.
The crude was purified by silica chromatography using hexanes as an
eluent to obtain bromopentacene derivative \textbf{4} as a deep blue solid in 16\% yield.

\textsuperscript{1}H-NMR (500 MHz, CDCl\textsubscript{3}, $\delta$ ppm): 9.72
(s, 1H), 9.41-9.33 (m, 3H), 8.04-7.96 (m, 3H), 7.77-7.75 (m, 1H),
7.48-7.45 (2H), 7.29-7.25 (m, 1H) and 1.48-1.38 (m, 42H).

\textsuperscript{13}C-NMR (125 MHz, CDCl\textsubscript{3}, $\delta$ ppm):
132.8, 132.5, 132.4, 130.94, 130.9, 130.8, 132.75, 130.72, 129.8, 128.9,
128.7, 128.66, 127.3, 126.9, 126.6, 126.3, 126.3, 126.2, 125.8, 123.4,
119.3, 118.3, 108.0, 107.5, 104.5, 104.3, 19.1, 19.0 and 11.7.

MS (ASAP): Calculated: 717.2947; Observed: 717.2939.

\subsubsection{Molecule 5}
\begin{figure}[h]
\includegraphics[width=.3\textwidth]{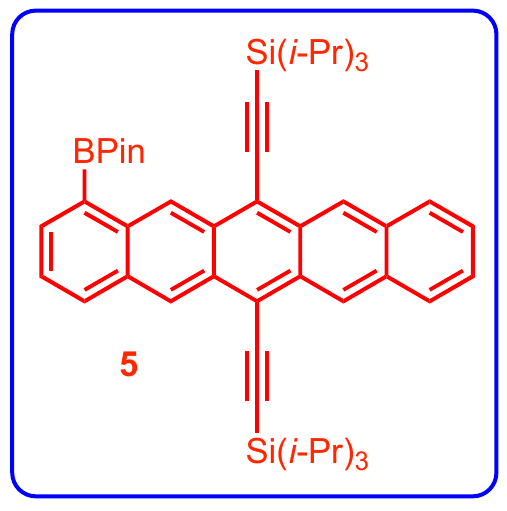}
\end{figure}
To a dry round bottomed flask was added \textbf{4} (4.0 g, 5.57 mmol),
Pd(dppf)Cl\textsubscript{2}$\cdot$DCM (203 mg, 0.25 mmol), KOAc (1.91 g, 19.5
mmol), and bis(pinacolato)diboron (2.82g, 11.1 mmol). Sequential vacuum
and argon were used to degas the mixture followed by the addition of
degassed 1, 4 dioxane (70 mL). The mixture was heated to 80 \textbf{$^\circ$}C
and maintained for 12 h in the dark. After the reaction, the mixture was
cooled to rt and the solvent was removed under reduced pressure. The
crude was partitioned between DCM (250 mL) and water (200 mL). The
organic layer was separated, washed with water (2 $\times$— 200 mL), dried over
\emph{anhyd.} Na\textsubscript{2}SO\textsubscript{4}, filtered and the
solvent was removed under reduced pressure to get the crude product. The
crude was purified by silica chromatography using mixtures of
hexanes/DCM as an eluent to obtain BPin pentacene derivative \textbf{5} as a deep
blue solid in 80\% yield.

\textsuperscript{1}H-NMR (400 MHz, CDCl\textsubscript{3}, $\delta$ ppm): 9.93
(s, 1H), 9.52 (s, 1H), 9.37 (s, 2H), 8.12-8.08 (m, 2H), 8.04-8.00 (m,
2H), 7.47-7.44 (m, 3H), 1.54 (s, 12H) and 1.44-1.40 (m, 42H).

\textsuperscript{13}C-NMR (100 MHz, CDCl\textsubscript{3}, $\delta$ ppm):
137.4, 134.5, 132.6, 132.4, 132.3, 132.2, 131.6, 130.7, 130.6, 130.4,
128.6, 127.3, 126.8, 126.7, 126.2, 125.96, 125.1, 118.9, 118.3, 107.6,
107.0, 105.5, 104.9, 83.9, 24.9, 19.2, 19.0, 12.0 and 11.7.

MS (ASAP): Calculated: 765.4694; Observed: 765.4688


\subsection{Synthesis of bipentacenes:} 
To a 20 mL sealed tube was added
\textbf{4a/b} (72 mg, 0.1 mmol), \textbf{5a/b} (76 mg, 0.1 mmol),
Pd(dppf)Cl\textsubscript{2}$\cdot$DCM (4 mg, 0.005 mmol), and
K\textsubscript{2}CO\textsubscript{3} (240 mg, 1.7 mmol). Sequential
vacuum and argon were used to degas the mixture followed by the addition
of degassed H2O (1 mL) and THF (3 mL). The resulting solution was heated
to 70 \textbf{$^\circ$}C and maintained for 24 h in dark. After the reaction,
the solution was poured into a separatory funnel containing DCM (30 mL)
and water (30 mL). The organic layer was separated, dried over
\emph{anhyd.} Na\textsubscript{2}SO\textsubscript{4}, filtered and the
solvent was removed under reduced pressure to get the crude product. The
crude was purified by silica chromatography using mixtures of
hexanes/DCM as an eluent to obtain bis-pentacenes as solid.

\subsubsection{11BP}

\begin{figure}[h]
\includegraphics[width=.3\textwidth]{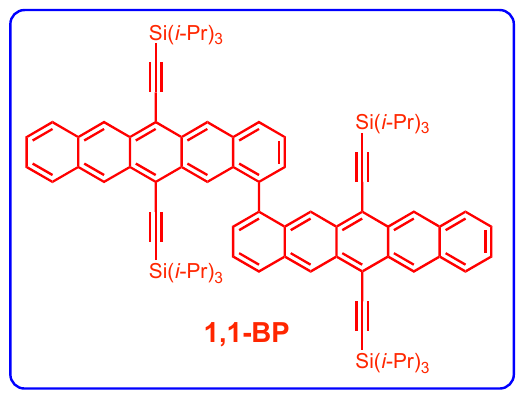}
\end{figure}
Yield:
84\% (Pale blue solid)

\textsuperscript{1}H-NMR (500 MHz, CDCl\textsubscript{3}, $\delta$ ppm): 9.51
(s, 1H), 9.31 (s, 1H), 9.15 (s, 1H), 9.07 (s, 1H), 8.15 (d, \emph{J} =
8.5 Hz, 1H), 7.97 (d, \emph{J} = 8.5 Hz, 1H), 7.85 (d, \emph{J} = 8.5
Hz, 1H), 7.61-7.58 (m, 1H), 7.52-7.51 (m, 1H), 7.41-7.34 (m, 2H),
1.496-1.44 (m, 21H), 0.87-0.86 (m, 9H) and 0.76-0.68 (m, 12H).

\textsuperscript{13}C-NMR (125 MHz, CDCl\textsubscript{3}, $\delta$ ppm):
139.3, 132.7, 132.3, 132.2, 132.1, 130.8, 130.6, 129.1, 128.7, 128.6,
128.2, 126.9, 126.5, 126.1, 125.9, 125.9, 125.8, 125.7, 118.9, 117.9,
107.4, 106.8, 104.9, 103.8, 19.1, 18.6, 18.5, 11.8 and 11.2.

MS (ASAP): Calculated: 1275.7450; Observed: 1275.7440

\subsubsection{22eBP}
\begin{figure}[h]
\includegraphics[width=.5\textwidth]{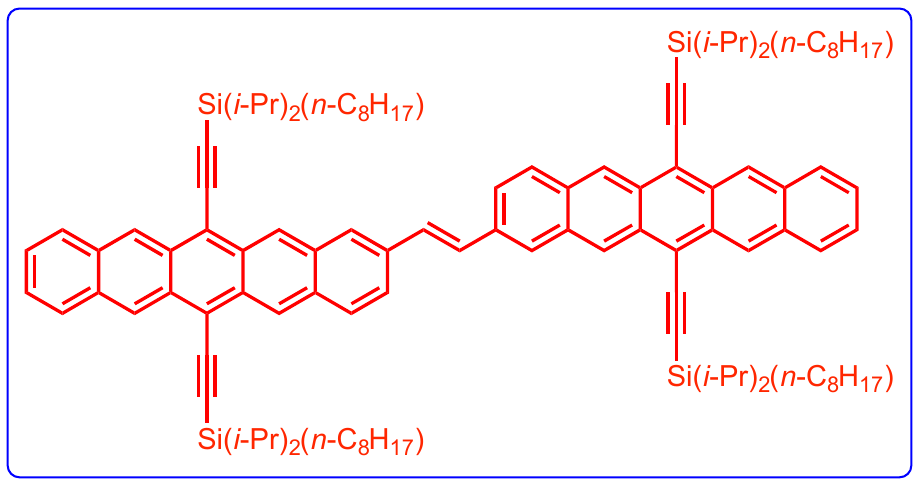}
\end{figure}
To
a 20 mL sealed tube was added \textbf{4c} (240 mg, 0.28 mmol),
Pd(dppf)Cl\textsubscript{2}$\cdot$DCM (12 mg, 0.015 mmol), and
trans-1,2-bis(tri-n-butylstannyl)ethylene \textbf{6} (66 mg, 0.11 mmol).
Sequential vacuum and argon were used to degas the mixture followed by
the addition of degassed THF (3 mL). The mixture was heated to reflux
and maintained for 5 h in the dark. After the reaction, the mixture was
cooled to rt and the solvent was removed under reduced pressure. The
crude reaction mixture was purified by silica chromatography using
mixtures of hexanes/DCM as an eluent to obtain product as a deep blue
solid with 80\% yield.

\textsuperscript{1}H-NMR (500 MHz, CDCl\textsubscript{3}, $\delta$ ppm):
9.32-9.23 (m, 8H), 8.02-7.97 (m, 8H), 7.86-7.84 (m, 2H), 7.50 (s, 2H),
7.49-7.42 (m, 4H), 1.85-1.77 (m, 8H), 1.58-1.54 (8H, overlap
H\textsubscript{2}O), 1.42-1.34 (m, 70H), 1.29-1.27 (m, 16H), 1.03-0.98
(m, 9H) and 0.89-0.81 (m, 13H)

\textsuperscript{13}C-NMR (125 MHz, CDCl\textsubscript{3}, $\delta$ ppm):
134.6, 132.4, 132.3, 132.2, 131.9, 131.0, 130.8, 130.7, 130.5, 129.1,
128.6, 128.2, 126.4, 126.3, 126.2, 126.0, 125.99, 125.91, 123.1, 118.4,
118.2, 107.4, 107.3, 104.6, 104.5, 34.1, 34.07, 32.0, 29.6, 29.5, 29.44,
29.41, 25.0, 24.99, 22.7, 18.8, 18.77, 18.72, 18.5, 18.49, 14.1, 14.08,
12.3, 12.2, 10.6 and 10.5.

MS (ESI): Calculated: 1582.0712; Observed: 1582.0714

\clearpage

\section{Co-ordinates for electronic structure calculations}
\label{sec:coord}
All coordinates are given in \AA ngstr\"oms in GAMESS notation.
\subsection{TIPS-pentacene}

Optimized in the $D_{2h}$ point group.
\begin{verbatim}
   ATOM   CHARGE       X              Y              Z
 ------------------------------------------------------------
 C           6.0   0.0000000000   4.0308083610   0.0000000000
 C           6.0   0.0000000000  -4.0308083610   0.0000000000
 C           6.0   0.0000000000   2.8480468252   0.0000000000
 C           6.0   0.0000000000  -2.8480468252   0.0000000000
 C           6.0   0.0000000000   1.4087548890   0.0000000000
 C           6.0  -0.0000000000  -1.4087548890   0.0000000000
 C           6.0  -1.2241740825   0.7138331268   0.0000000000
 C           6.0   1.2241740825   0.7138331268   0.0000000000
 C           6.0  -1.2241740825  -0.7138331268   0.0000000000
 C           6.0   1.2241740825  -0.7138331268   0.0000000000
 C           6.0  -2.4645223735   1.3937263190   0.0000000000
 C           6.0   2.4645223735   1.3937263190   0.0000000000
 C           6.0  -2.4645223735  -1.3937263190   0.0000000000
 C           6.0   2.4645223735  -1.3937263190   0.0000000000
 C           6.0  -3.6641994773   0.7090627815   0.0000000000
 C           6.0   3.6641994773   0.7090627815   0.0000000000
 C           6.0  -3.6641994773  -0.7090627815   0.0000000000
 C           6.0   3.6641994773  -0.7090627815   0.0000000000
 C           6.0  -6.0812759032   0.7175412661   0.0000000000
 C           6.0   6.0812759032   0.7175412661   0.0000000000
 C           6.0  -6.0812759032  -0.7175412661   0.0000000000
 C           6.0   6.0812759032  -0.7175412661   0.0000000000
 C           6.0  -4.9258450649   1.3986591718   0.0000000000
 C           6.0   4.9258450649   1.3986591718   0.0000000000
 C           6.0  -4.9258450649  -1.3986591718   0.0000000000
 C           6.0   4.9258450649  -1.3986591718   0.0000000000
 H           1.0   0.0000000000   5.0839495280   0.0000000000
 H           1.0   0.0000000000  -5.0839495280   0.0000000000
 H           1.0  -2.4695833345   2.4631338945   0.0000000000
 H           1.0   2.4695833345   2.4631338945   0.0000000000
 H           1.0  -2.4695833345  -2.4631338945   0.0000000000
 H           1.0   2.4695833345  -2.4631338945   0.0000000000
 H           1.0  -4.9230541382   2.4715419620   0.0000000000
 H           1.0   4.9230541382   2.4715419620   0.0000000000
 H           1.0  -4.9230541382  -2.4715419620   0.0000000000
 H           1.0   4.9230541382  -2.4715419620   0.0000000000
 H           1.0   7.0179156581  -1.2400824719   0.0000000000
 H           1.0  -7.0179156581  -1.2400824719   0.0000000000
 H           1.0   7.0179156581   1.2400824719   0.0000000000
 H           1.0  -7.0179156581   1.2400824719   0.0000000000
  \end{verbatim}

\subsection{11BP}
Optimized in the $C_2$ point group. 
A number of low-energy structures were found with slightly different dihedral angles (80$^\circ$--100$^\circ$) between the two pentacenes, but which did not significantly alter the calculated spectra. We chose a structure with a dihedral of 81$^\circ$ which was supported by low-level calculations on 11BP with the TIPS groups.
\begin{verbatim}
 C           6.0  -5.0168112146   1.6786790667   0.0464831369
 C           6.0   5.0168112146  -1.6786790667   0.0464831369
 C           6.0  -3.7770489120  -0.8995243633  -0.0234142169
 C           6.0   3.7770489120   0.8995243633  -0.0234142169
 C           6.0  -5.6341113352   2.9610319681   0.0836332259
 C           6.0   5.6341113352  -2.9610319681   0.0836332259
 C           6.0  -3.1608849614  -2.1824592852  -0.0580493116
 C           6.0   3.1608849614   2.1824592852  -0.0580493116
 C           6.0  -6.1618397466   4.0524937895   0.1147753046
 C           6.0   6.1618397466  -4.0524937895   0.1147753046
 C           6.0  -2.6362554178  -3.2751541152  -0.0859442496
 C           6.0   2.6362554178   3.2751541152  -0.0859442496
 H           1.0  -6.6270642889   5.0117618599   0.1402879113
 H           1.0   6.6270642889  -5.0117618599   0.1402879113
 H           1.0  -2.1654838949  -4.2317995341  -0.1083126340
 H           1.0   2.1654838949   4.2317995341  -0.1083126340
 C           6.0  -5.5622687669   0.6571259705  -0.7810752597
 C           6.0   5.5622687669  -0.6571259705  -0.7810752597
 C           6.0  -3.8603762837   1.4291449469   0.8376328753
 C           6.0   3.8603762837  -1.4291449469   0.8376328753
 C           6.0  -4.9366736721  -0.6530391474  -0.8115082957
 C           6.0   4.9366736721   0.6530391474  -0.8115082957
 C           6.0  -3.2276382667   0.1245364571   0.7985422638
 C           6.0   3.2276382667  -0.1245364571   0.7985422638
 C           6.0  -6.7048631222   0.8805966134  -1.5805194842
 C           6.0   6.7048631222  -0.8805966134  -1.5805194842
 C           6.0  -3.2998609203   2.4214212570   1.6698352866
 C           6.0   3.2998609203  -2.4214212570   1.6698352866
 C           6.0  -5.5017716946  -1.6548483714  -1.6313085290
 C           6.0   5.5017716946   1.6548483714  -1.6313085290
 C           6.0  -2.0748711290  -0.0989087129   1.5847195629
 C           6.0   2.0748711290   0.0989087129   1.5847195629
 H           1.0  -7.1682027183   1.8623796989  -1.5602718633
 H           1.0   7.1682027183  -1.8623796989  -1.5602718633
 H           1.0  -3.7724288545   3.3985212045   1.7025429323
 H           1.0   3.7724288545  -3.3985212045   1.7025429323
 H           1.0  -5.0342209852  -2.6346700386  -1.6485433777
 H           1.0   5.0342209852   2.6346700386  -1.6485433777
 H           1.0  -1.6088469264  -1.0763983682   1.5451099309
 H           1.0   1.6088469264   1.0763983682   1.5451099309
 C           6.0  -7.2493988858  -0.1126690066  -2.3876427575
 C           6.0   7.2493988858   0.1126690066  -2.3876427575
 C           6.0  -2.1669125893   2.1903869510   2.4432341397
 C           6.0   2.1669125893  -2.1903869510   2.4432341397
 C           6.0  -6.6303294657  -1.4242279454  -2.4106306834
 C           6.0   6.6303294657   1.4242279454  -2.4106306834
 C           6.0  -1.5280677528   0.8865209381   2.4013913396
 C           6.0   1.5280677528  -0.8865209381   2.4013913396
 C           6.0  -8.4084149891   0.1132809950  -3.2031695924
 C           6.0   8.4084149891  -0.1132809950  -3.2031695924
 C           6.0  -1.6120311759   3.2120194516   3.2823688117
 C           6.0   1.6120311759  -3.2120194516   3.2823688117
 C           6.0  -7.2071942329  -2.4409706094  -3.2433249054
 C           6.0   7.2071942329   2.4409706094  -3.2433249054
 C           6.0  -0.3437935893   0.6649544380   3.2076966897
 C           6.0   0.3437935893  -0.6649544380   3.2076966897
 H           1.0  -8.8683908544   1.0986165748  -3.1862532429
 H           1.0   8.8683908544  -1.0986165748  -3.1862532429
 H           1.0  -2.1010646062   4.1830056111   3.3088108062
 H           1.0   2.1010646062  -4.1830056111   3.3088108062
 H           1.0  -6.7426838926  -3.4242233517  -3.2556394540
 H           1.0   6.7426838926   3.4242233517  -3.2556394540
 C           6.0  -8.9228223218  -0.8846949932  -3.9831660604
 C           6.0   8.9228223218   0.8846949932  -3.9831660604
 C           6.0  -0.4946723195   2.9676969390   4.0285519904
 C           6.0   0.4946723195  -2.9676969390   4.0285519904
 C           6.0  -8.3140535576  -2.1814329492  -4.0025104129
 C           6.0   8.3140535576   2.1814329492  -4.0025104129
 C           6.0   0.1409618000   1.6875984915   3.9854886973
 C           6.0  -0.1409618000  -1.6875984915   3.9854886973
 H           1.0  -9.8016592260  -0.7011761381  -4.5958530875
 H           1.0   9.8016592260   0.7011761381  -4.5958530875
 H           1.0  -0.0768503612   3.7433357089   4.6650280256
 H           1.0   0.0768503612  -3.7433357089   4.6650280256
 H           1.0  -8.7419485671  -2.9596339230  -4.6292654463
 H           1.0   8.7419485671   2.9596339230  -4.6292654463
 H           1.0   1.0306306985   1.5196266204   4.5868899340
 H           1.0  -1.0306306985  -1.5196266204   4.5868899340
\end{verbatim}

\subsection{22BP}
Structure taken from Ref.~\cite{fue16a}.

\subsection{22eBP}
Optimized in the $C_{2h}$ point group.
\begin{verbatim}
C   6.0     -7.92668247    -1.26516759     0.00000000
C   6.0      7.92667723     1.26516676     0.00000000
C   6.0     -8.31315231     1.56966603     0.00000000
C   6.0      8.31314659    -1.56966686     0.00000000
C   6.0     -7.73448944    -2.67577052     0.00000000
C   6.0      7.73448372     2.67576981     0.00000000
C   6.0     -8.50554085     2.98072267     0.00000000
C   6.0      8.50553513    -2.98072338     0.00000000
C   6.0     -7.57011127    -3.87732530     0.00000000
C   6.0      7.57005262     3.87731624     0.00000000
C   6.0     -8.67098522     4.18200779     0.00000000
C   6.0      8.67100811    -4.18200445     0.00000000
H   1.0     -7.42592239    -4.93404913     0.00000000
H   1.0      7.42644835     4.93412924     0.00000000
H   1.0     -8.81589413     5.23868799     0.00000000
H   1.0      8.81561184    -5.23873091     0.00000000
C   6.0     -9.24816132    -0.73380011     0.00000000
C   6.0      9.24815559     0.73379898     0.00000000
C   6.0     -6.79677439    -0.40012172     0.00000000
C   6.0      6.79676867     0.40012118     0.00000000
C   6.0     -9.44531250     0.70449984     0.00000000
C   6.0      9.44530773    -0.70450091     0.00000000
C   6.0     -6.99528408     1.03862977     0.00000000
C   6.0      6.99527884    -1.03863037     0.00000000
C   6.0    -10.38371086    -1.57286096     0.00000000
C   6.0     10.38370514     1.57286048     0.00000000
C   6.0     -5.47686577    -0.90274304     0.00000000
C   6.0      5.47686005     0.90274173     0.00000000
C   6.0    -10.76369190     1.20770228     0.00000000
C   6.0     10.76368713    -1.20770264     0.00000000
C   6.0     -5.85674047     1.87935269     0.00000000
C   6.0      5.85673523    -1.87935376     0.00000000
H   1.0    -10.23241329    -2.64811635     0.00000000
H   1.0     10.23240566     2.64811540     0.00000000
H   1.0     -5.33027649    -1.97862923     0.00000000
H   1.0      5.33026886     1.97862792     0.00000000
H   1.0    -10.90994263     2.28368711     0.00000000
H   1.0     10.90993786    -2.28368759     0.00000000
H   1.0     -6.00853205     2.95449662     0.00000000
H   1.0      6.00852776    -2.95449758     0.00000000
C   6.0    -11.67914391    -1.06567049     0.00000000
C   6.0     11.67913818     1.06567049     0.00000000
C   6.0     -4.36550093    -0.06438924     0.00000000
C   6.0      4.36549520     0.06438758     0.00000000
C   6.0    -11.87567043     0.37100893     0.00000000
C   6.0     11.87566471    -0.37100887     0.00000000
C   6.0     -4.56643677     1.37154436     0.00000000
C   6.0      4.56643152    -1.37154591     0.00000000
C   6.0    -12.83485413    -1.91612911     0.00000000
C   6.0     12.83484745     1.91612971     0.00000000
C   6.0     -3.02918434    -0.56868756     0.00000000
C   6.0      3.02917838     0.56868529     0.00000000
C   6.0    -13.21658802     0.88056803     0.00000000
C   6.0     13.21658325    -0.88056731     0.00000000
C   6.0     -3.40303636     2.21517420     0.00000000
C   6.0      3.40303159    -2.21517611     0.00000000
H   1.0    -12.68388176    -2.99316859     0.00000000
H   1.0     12.68387508     2.99316907     0.00000000
H   1.0     -2.88808918    -1.64788556     0.00000000
H   1.0      2.88808250     1.64788330     0.00000000
H   1.0    -13.36110401     1.95852780     0.00000000
H   1.0     13.36109924    -1.95852697     0.00000000
H   1.0     -3.54865170     3.29301834     0.00000000
H   1.0      3.54864740    -3.29302025     0.00000000
C   6.0    -14.09543991    -1.38647938     0.00000000
C   6.0     14.09543324     1.38648045     0.00000000
C   6.0     -1.92513561     0.26085106     0.00000000
C   6.0      1.92513001    -0.26085377     0.00000000
C   6.0    -14.28925800     0.03266053     0.00000000
C   6.0     14.28925323    -0.03265931     0.00000000
C   6.0     -2.14609814     1.69066441     0.00000000
C   6.0      2.14609313    -1.69066703     0.00000000
H   1.0    -14.96322536    -2.04089022     0.00000000
H   1.0     14.96321869     2.04089165     0.00000000
H   1.0    -15.30041599     0.43142483     0.00000000
H   1.0     15.30041027    -0.43142313     0.00000000
H   1.0     -1.29143643     2.35977221     0.00000000
H   1.0      1.29143167    -2.35977530     0.00000000
C   6.0     -0.59145021    -0.32948795     0.00000000
C   6.0      0.59144425     0.32948473     0.00000000
H   1.0     -0.58765703    -1.41819751     0.00000000
H   1.0      0.58765107     1.41819429     0.00000000
\end{verbatim}

\subsection{66yBP}
This structure was optimized in the $D_{2h}$ point group. Test calculations were performed on a twisted structure ($D_2$ symmetry) and found it of very slightly higher energy.
\begin{verbatim}
 C           6.0   8.8126464405   0.0000000000   0.0000000000
 C           6.0  -8.8126464405   0.0000000000   0.0000000000
 C           6.0   0.6754618538   0.0000000000   0.0000000000
 C           6.0  -0.6754618538   0.0000000000   0.0000000000
 C           6.0   7.5997592132   0.0000000000   0.0000000000
 C           6.0  -7.5997592132   0.0000000000   0.0000000000
 C           6.0   1.9039435613   0.0000000000   0.0000000000
 C           6.0  -1.9039435613   0.0000000000   0.0000000000
 C           6.0   6.1769124108   0.0000000000   0.0000000000
 C           6.0  -6.1769124108   0.0000000000   0.0000000000
 C           6.0   3.3101132040   0.0000000000   0.0000000000
 C           6.0  -3.3101132040   0.0000000000   0.0000000000
 C           6.0   5.4725886874  -1.2381061438   0.0000000000
 C           6.0  -5.4725886874  -1.2381061438   0.0000000000
 C           6.0   5.4725886874   1.2381061438   0.0000000000
 C           6.0  -5.4725886874   1.2381061438   0.0000000000
 C           6.0   4.0214784718  -1.2414960540   0.0000000000
 C           6.0  -4.0214784718  -1.2414960540   0.0000000000
 C           6.0   4.0214784718   1.2414960540   0.0000000000
 C           6.0  -4.0214784718   1.2414960540   0.0000000000
 C           6.0   6.1528546455  -2.4750915850   0.0000000000
 C           6.0  -6.1528546455  -2.4750915850   0.0000000000
 C           6.0   6.1528546455   2.4750915850   0.0000000000
 C           6.0  -6.1528546455   2.4750915850   0.0000000000
 C           6.0   3.3482476722  -2.4788662615   0.0000000000
 C           6.0  -3.3482476722  -2.4788662615   0.0000000000
 C           6.0   3.3482476722   2.4788662615   0.0000000000
 C           6.0  -3.3482476722   2.4788662615   0.0000000000
 C           6.0   5.4778763675  -3.6918197777   0.0000000000
 C           6.0  -5.4778763675  -3.6918197777   0.0000000000
 C           6.0   5.4778763675   3.6918197777   0.0000000000
 C           6.0  -5.4778763675   3.6918197777   0.0000000000
 C           6.0   4.0282544584  -3.6942743126   0.0000000000
 C           6.0  -4.0282544584  -3.6942743126   0.0000000000
 C           6.0   4.0282544584   3.6942743126   0.0000000000
 C           6.0  -4.0282544584   3.6942743126   0.0000000000
 C           6.0   5.4735166467  -6.1287491908   0.0000000000
 C           6.0  -5.4735166467  -6.1287491908   0.0000000000
 C           6.0   5.4735166467   6.1287491908   0.0000000000
 C           6.0  -5.4735166467   6.1287491908   0.0000000000
 C           6.0   4.0414062574  -6.1309006058   0.0000000000
 C           6.0  -4.0414062574  -6.1309006058   0.0000000000
 C           6.0   4.0414062574   6.1309006058   0.0000000000
 C           6.0  -4.0414062574   6.1309006058   0.0000000000
 C           6.0   6.1666750376  -4.9500376145   0.0000000000
 C           6.0  -6.1666750376  -4.9500376145   0.0000000000
 C           6.0   6.1666750376   4.9500376145   0.0000000000
 C           6.0  -6.1666750376   4.9500376145   0.0000000000
 C           6.0   3.3442700178  -4.9541008366   0.0000000000
 C           6.0  -3.3442700178  -4.9541008366   0.0000000000
 C           6.0   3.3442700178   4.9541008366   0.0000000000
 C           6.0  -3.3442700178   4.9541008366   0.0000000000
 H           1.0   9.8791853511   0.0000000000   0.0000000000
 H           1.0  -9.8791853511   0.0000000000   0.0000000000
 H           1.0   7.2386447019  -2.4700717770   0.0000000000
 H           1.0  -7.2386447019  -2.4700717770   0.0000000000
 H           1.0   7.2386447019   2.4700717770   0.0000000000
 H           1.0  -7.2386447019   2.4700717770   0.0000000000
 H           1.0   2.2621663549  -2.4804311889   0.0000000000
 H           1.0  -2.2621663549  -2.4804311889   0.0000000000
 H           1.0   2.2621663549   2.4804311889   0.0000000000
 H           1.0  -2.2621663549   2.4804311889   0.0000000000
 H           1.0   7.2542661597  -4.9448454951   0.0000000000
 H           1.0  -7.2542661597  -4.9448454951   0.0000000000
 H           1.0   7.2542661597   4.9448454951   0.0000000000
 H           1.0  -7.2542661597   4.9448454951   0.0000000000
 H           1.0   2.2566699811  -4.9524860584   0.0000000000
 H           1.0  -2.2566699811  -4.9524860584   0.0000000000
 H           1.0   2.2566699811   4.9524860584   0.0000000000
 H           1.0  -2.2566699811   4.9524860584   0.0000000000
 H           1.0   3.5118557631   7.0800143782   0.0000000000
 H           1.0  -3.5118557631   7.0800143782   0.0000000000
 H           1.0   3.5118557631  -7.0800143782   0.0000000000
 H           1.0  -3.5118557631  -7.0800143782   0.0000000000
 H           1.0   6.0058072668   7.0763213807   0.0000000000
 H           1.0  -6.0058072668   7.0763213807   0.0000000000
 H           1.0   6.0058072668  -7.0763213807   0.0000000000
 H           1.0  -6.0058072668  -7.0763213807   0.0000000000
\end{verbatim}

\clearpage

\section{NMR data}

\textsuperscript{1}H-NMR (500 MHz, CDCl\textsubscript{3}, $\delta$ ppm): 9.72
(s, 1H), 9.41-9.33 (m, 3H), 8.04-7.96 (m, 3H), 7.77-7.75 (m, 1H),
7.48-7.45 (2H), 7.29-7.25 (m, 1H) and 1.48-1.38 (m, 42H).

\begin{figure}[h]
\includegraphics[width=.38\textwidth]{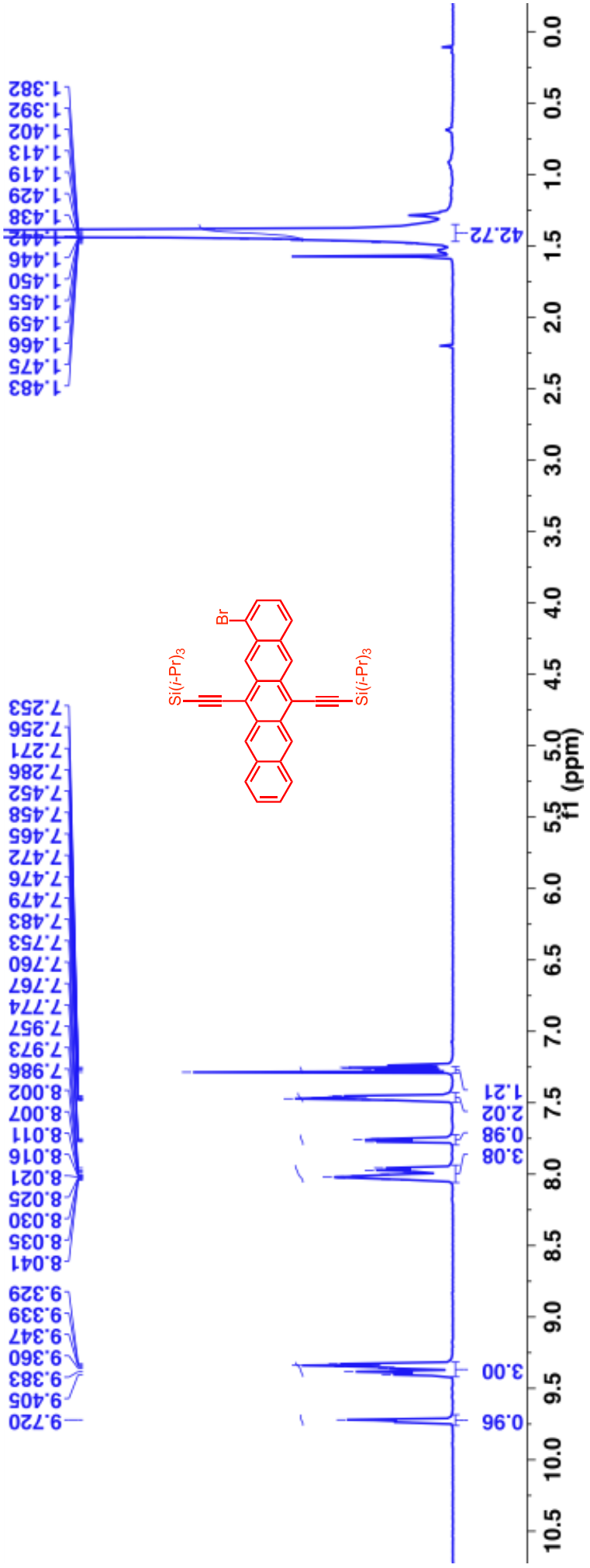}
\end{figure} \clearpage

\textsuperscript{13}C-NMR (125 MHz, CDCl\textsubscript{3}, $\delta$ ppm):
132.8, 132.5, 132.4, 130.94, 130.9, 130.8, 132.75, 130.72, 129.8, 128.9,
128.7, 128.66, 127.3, 126.9, 126.6, 126.3, 126.3, 126.2, 125.8, 123.4,
119.3, 118.3, 108.0, 107.5, 104.5, 104.3, 19.1, 19.0 and 11.7.
\begin{figure}[h]
\includegraphics[width=.38\textwidth]{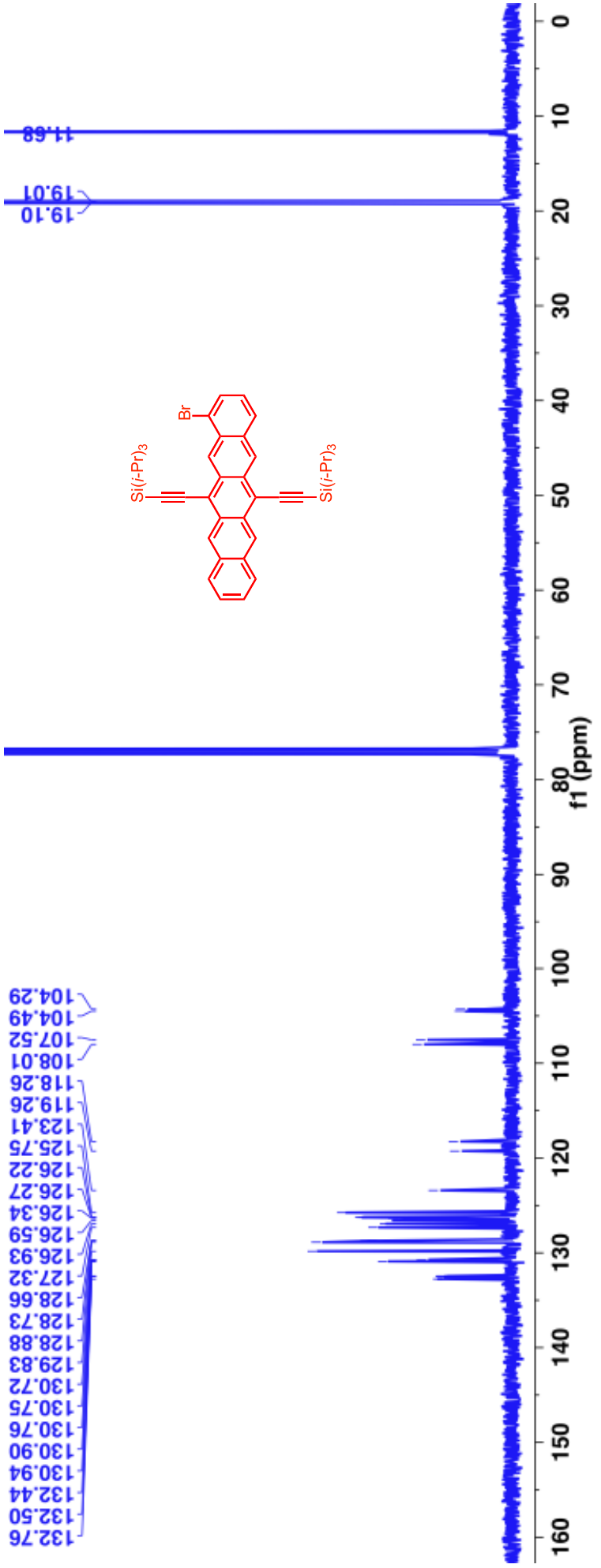}
\end{figure} \clearpage

\textsuperscript{1}H-NMR (400 MHz, CDCl\textsubscript{3}, $\delta$ ppm): 9.93
(s, 1H), 9.52 (s, 1H), 9.37 (s, 2H), 8.12-8.08 (m, 2H), 8.04-8.00 (m,
2H), 7.47-7.44 (m, 3H), 1.54 (s, 12H) and 1.44-1.40 (m, 42H).
\begin{figure}[h]
\includegraphics[width=.38\textwidth]{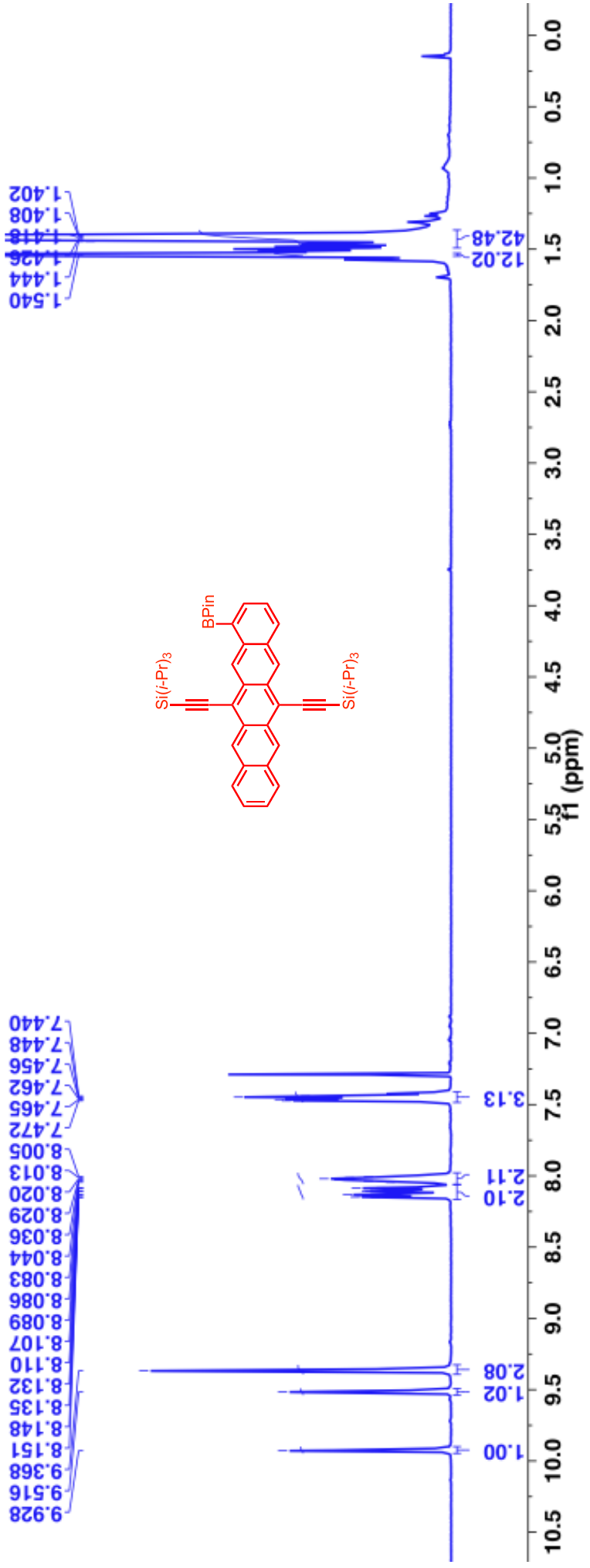}
\end{figure} \clearpage

\textsuperscript{13}C-NMR (100 MHz, CDCl\textsubscript{3}, $\delta$ ppm):
137.4, 134.5, 132.6, 132.4, 132.3, 132.2, 131.6, 130.7, 130.6, 130.4,
128.6, 127.3, 126.8, 126.7, 126.2, 125.96, 125.1, 118.9, 118.3, 107.6,
107.0, 105.5, 104.9, 83.9, 24.9, 19.2, 19.0, 12.0 and 11.7.
\begin{figure}[h]
\includegraphics[width=.38\textwidth]{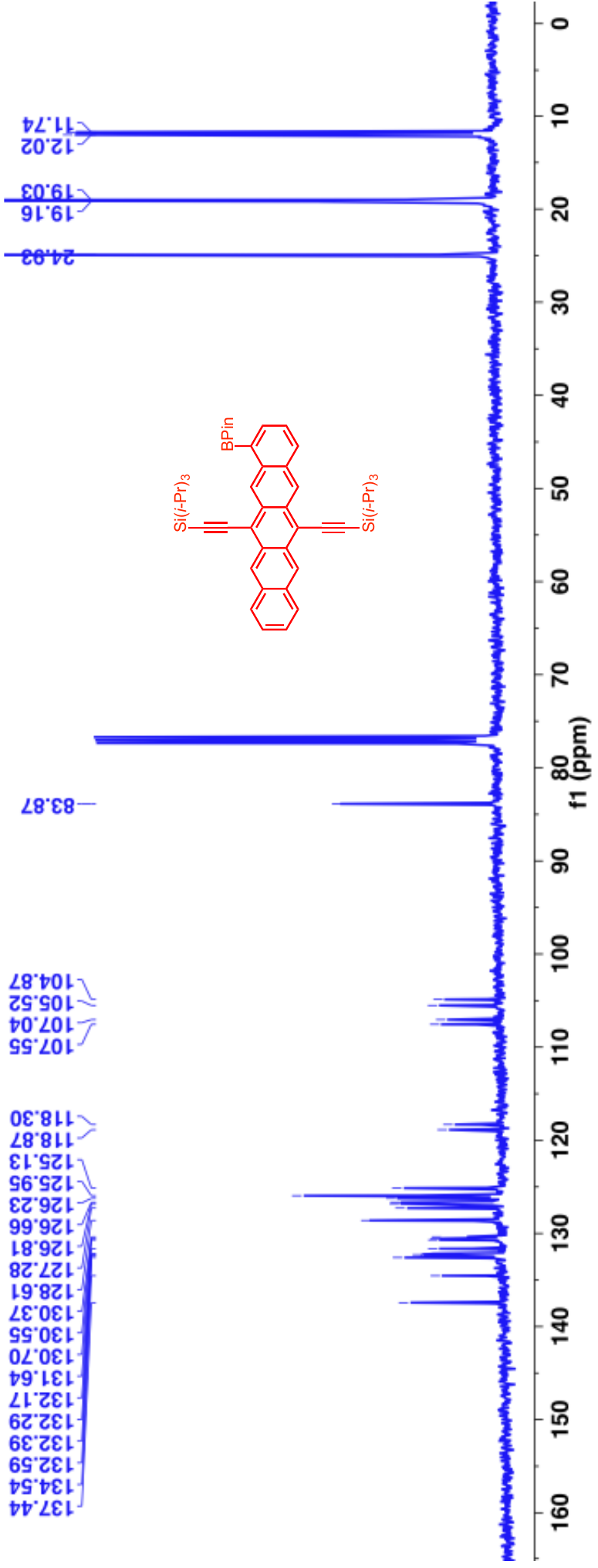}
\end{figure} \clearpage

\textsuperscript{1}H-NMR (500 MHz, CDCl\textsubscript{3}, $\delta$ ppm): 9.51
(s, 1H), 9.31 (s, 1H), 9.15 (s, 1H), 9.07 (s, 1H), 8.15 (d, \emph{J} =
8.5 Hz, 1H), 7.97 (d, \emph{J} = 8.5 Hz, 1H), 7.85 (d, \emph{J} = 8.5
Hz, 1H), 7.61-7.58 (m, 1H), 7.52-7.51 (m, 1H), 7.41-7.34 (m, 2H),
1.496-1.44 (m, 21H), 0.87-0.86 (m, 9H) and 0.76-0.68 (m, 12H).
\begin{figure}[h]
\includegraphics[width=.37\textwidth]{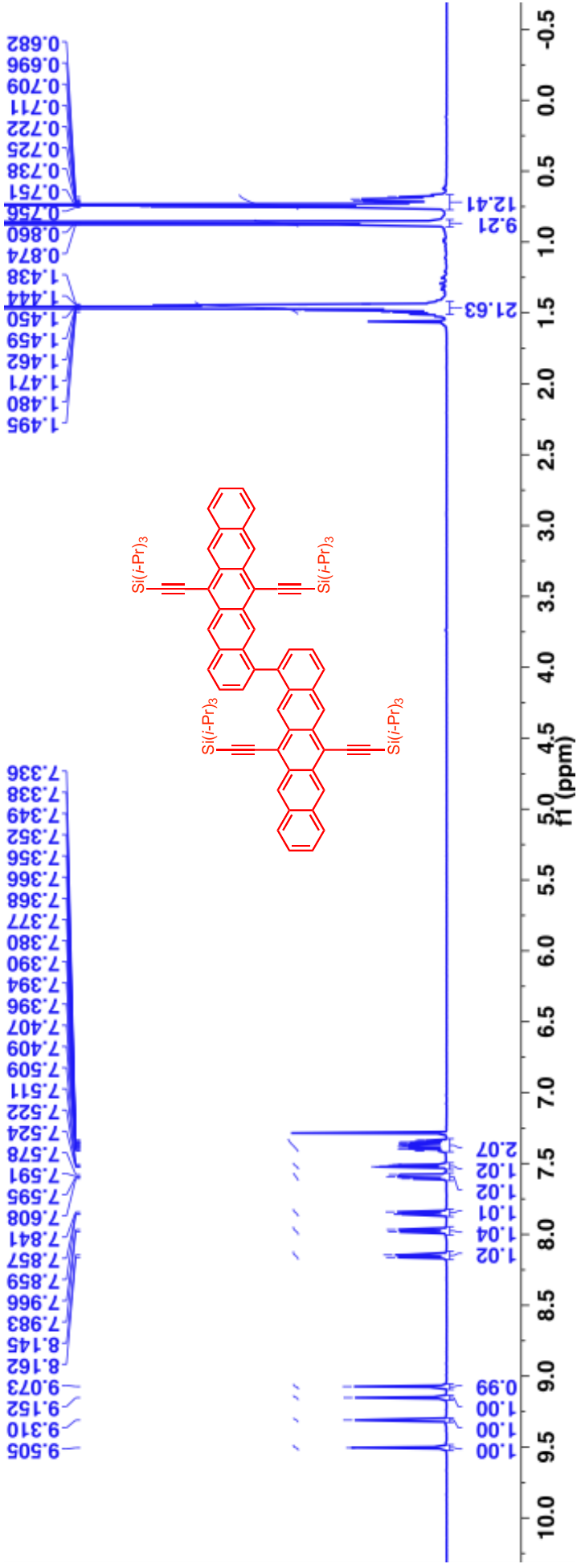}
\end{figure} \clearpage

\textsuperscript{13}C-NMR (125 MHz, CDCl\textsubscript{3}, $\delta$ ppm):
139.3, 132.7, 132.3, 132.2, 132.1, 130.8, 130.6, 129.1, 128.7, 128.6,
128.2, 126.9, 126.5, 126.1, 125.9, 125.9, 125.8, 125.7, 118.9, 117.9,
107.4, 106.8, 104.9, 103.8, 19.1, 18.6, 18.5, 11.8 and 11.2.
\begin{figure}[h]
\includegraphics[width=.38\textwidth]{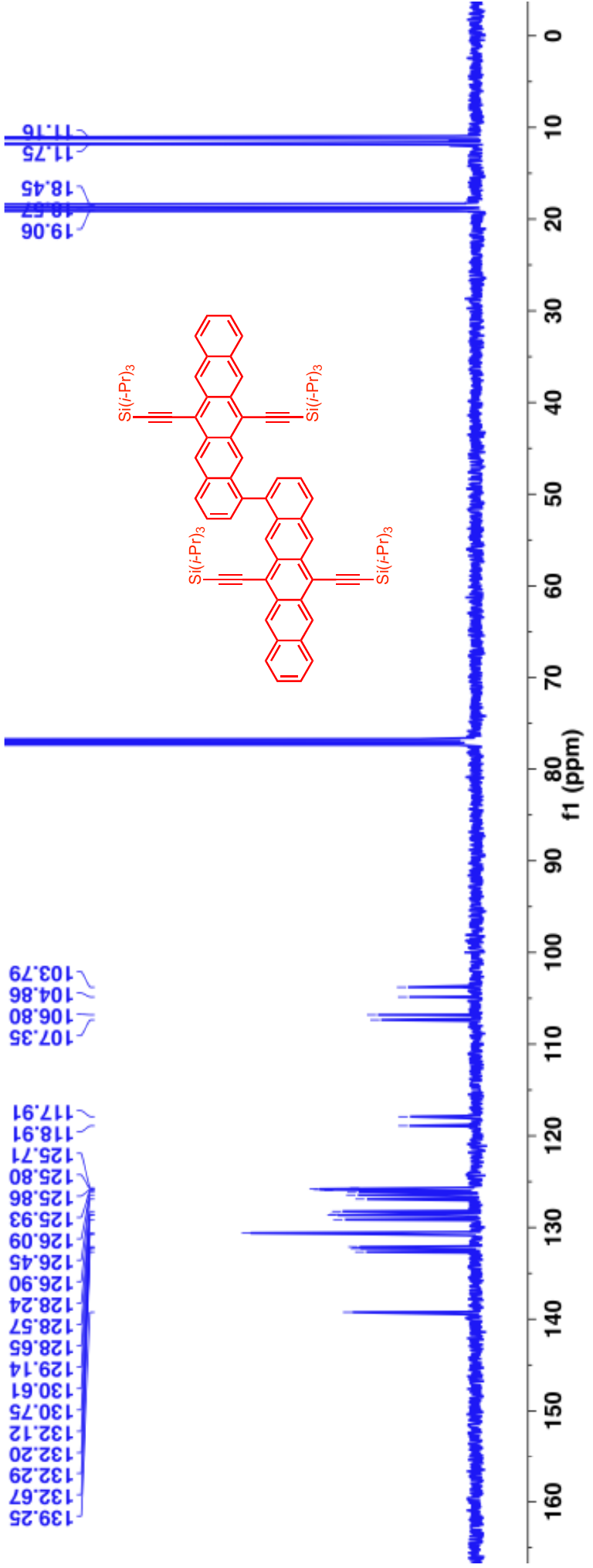}
\end{figure} \clearpage

\textsuperscript{1}H-NMR (500 MHz, CDCl\textsubscript{3}, $\delta$ ppm):
9.32-9.23 (m, 8H), 8.02-7.97 (m, 8H), 7.86-7.84 (m, 2H), 7.50 (s, 2H),
7.49-7.42 (m, 4H), 1.85-1.77 (m, 8H), 1.58-1.54 (8H, overlap
H\textsubscript{2}O), 1.42-1.34 (m, 70H), 1.29-1.27 (m, 16H), 1.03-0.98
(m, 9H) and 0.89-0.81 (m, 13H)
\begin{figure}[h]
\includegraphics[width=.38\textwidth]{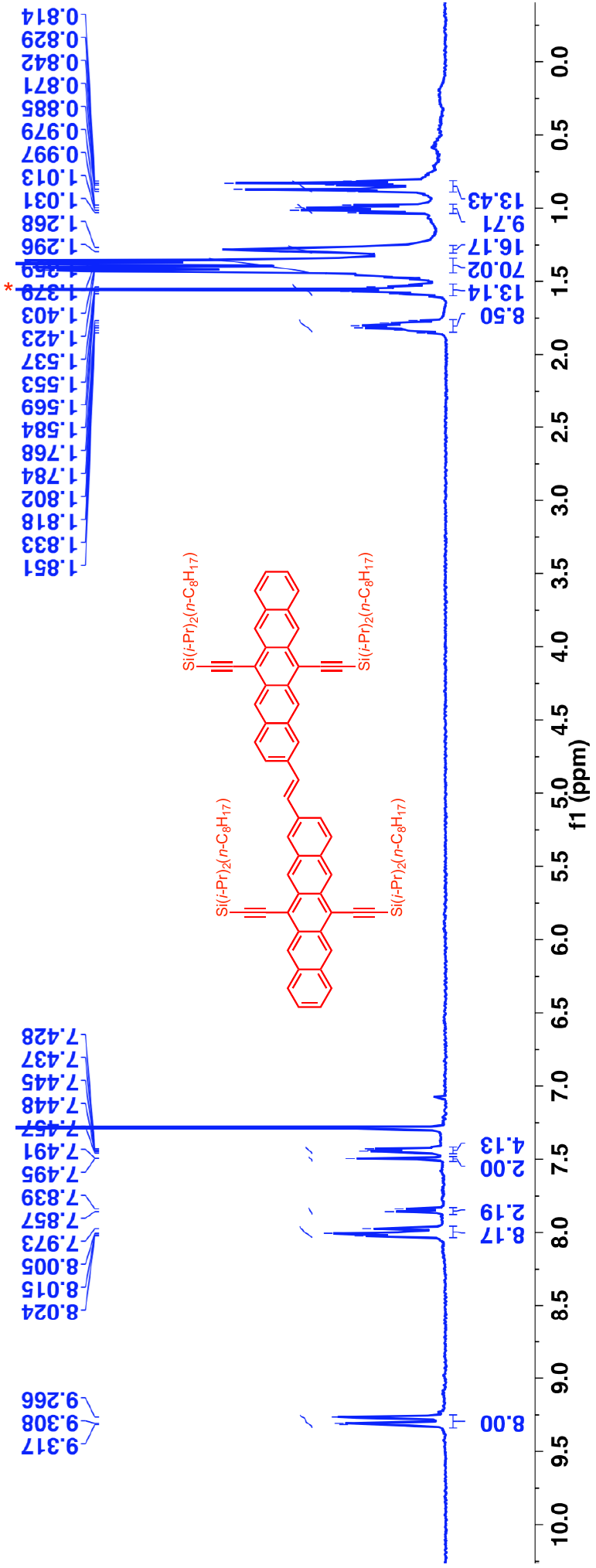}
\end{figure} \clearpage

\textsuperscript{13}C-NMR (125 MHz, CDCl\textsubscript{3}, $\delta$ ppm):
134.6, 132.4, 132.3, 132.2, 131.9, 131.0, 130.8, 130.7, 130.5, 129.1,
128.6, 128.2, 126.4, 126.3, 126.2, 126.0, 125.99, 125.91, 123.1, 118.4,
118.2, 107.4, 107.3, 104.6, 104.5, 34.1, 34.07, 32.0, 29.6, 29.5, 29.44,
29.41, 25.0, 24.99, 22.7, 18.8, 18.77, 18.72, 18.5, 18.49, 14.1, 14.08,
12.3, 12.2, 10.6 and 10.5.
\begin{figure}[h]
\includegraphics[width=.38\textwidth]{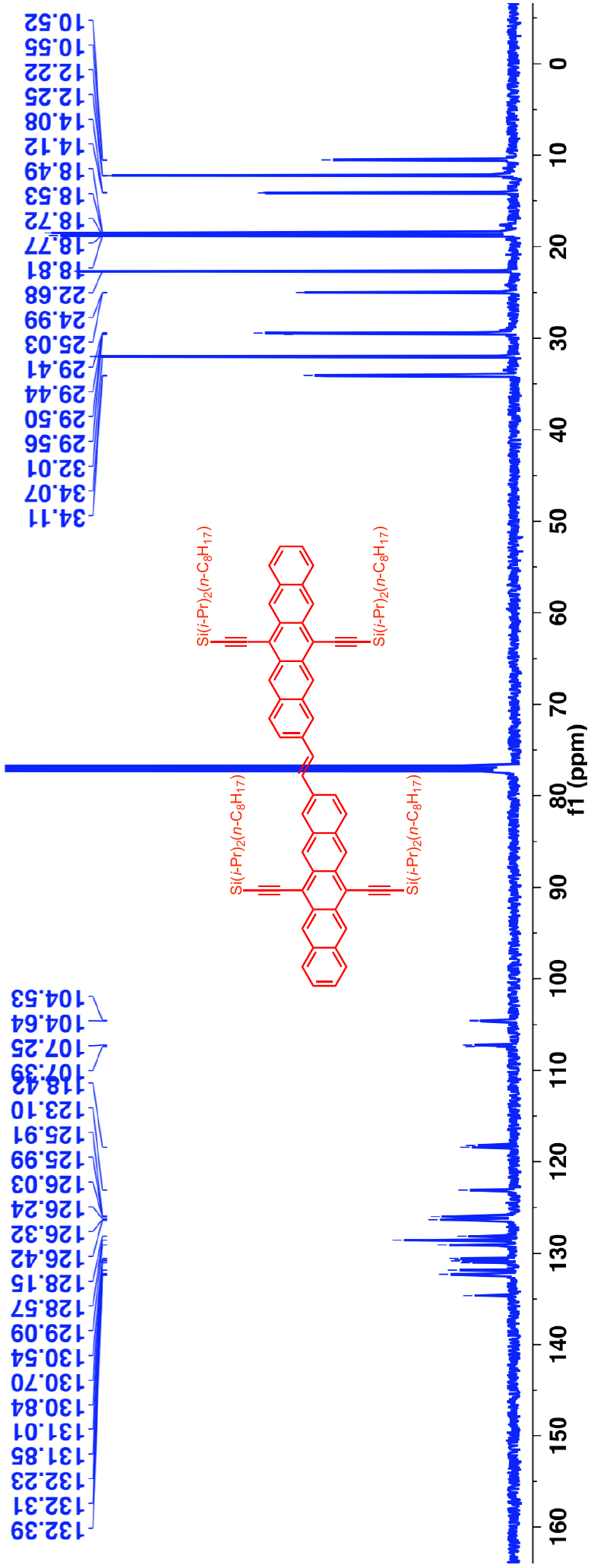}
\end{figure}

\bibliographystyle{biochem}
\bibliography{refbig}